\documentclass[12pt,letterpaper]{article}

\usepackage{epsfig}
\usepackage{amsmath}
\usepackage{amssymb}
\usepackage{amsthm}
\usepackage{indentfirst}
\usepackage{xspace}
\usepackage{multirow}
\usepackage{hyperref}
\usepackage{xcolor}
\definecolor{darkred}{rgb}{0.5,0.15,0.15}
\hypersetup{colorlinks=true,urlcolor=darkred,linkcolor=darkred,citecolor=darkred}
\usepackage{verbatim}
\usepackage[letterpaper,margin=1in,headheight=15pt]{geometry}
\usepackage{mathpazo}
\usepackage{authblk}
\usepackage{tikz-cd}
\usepackage{capt-of}

\usepackage{tikz}
\usetikzlibrary{calc}   
\usetikzlibrary{topaths}
\usetikzlibrary{decorations}
\usetikzlibrary{decorations.pathmorphing}

\numberwithin{equation}{section}

\newcommand{\fX}{{\mathfrak X}}

\newcommand{\cC}{\ensuremath{\mathcal C}}

\newcommand{\cS}{\ensuremath{\mathcal S}}

\newcommand{\cH}{\ensuremath{\mathcal H}}

\newcommand{\cW}{\ensuremath{\mathcal W}}

\newcommand{\bM}{\ensuremath{\mathbf M}}
\newcommand{\bF}{\ensuremath{\mathbf F}}
\newcommand{\bS}{\ensuremath{\mathbf S}}

\newcommand{\R}{\ensuremath{\mathbb R}}
\newcommand{\C}{\ensuremath{\mathbb C}}
\newcommand{\PP}{\ensuremath{\mathbb P}}
\newcommand{\Z}{\ensuremath{\mathbb Z}}

\newcommand{\bbS}{\ensuremath{\mathbb S}}
\newcommand{\bbL}{\ensuremath{\mathbb L}}

\newcommand{\half}{\ensuremath{\frac{1}{2}}}

\newcommand{\N}{{\mathcal N}}

\newcommand{\ored}{\Omega_{\mathrm{red}}}
\newcommand{\bored}{\bom_{\mathrm{red}}}
\newcommand{\bom}{{\mathbf \Omega}}
\newcommand{\bL}{{\mathbf L}}
\newcommand{\bal}{{\boldsymbol {\alpha}}}

\newcommand{\I}{{\mathrm i}}
\newcommand{\E}{{\mathrm e}}
\newcommand{\de}{\mathrm{d}}

\newcommand{\abs}[1]{\lvert#1\rvert}

\newcommand{\fro}{{\overline{\underline{\Omega}}}}

\newcommand{\ti}[1]{\textit{#1}}

\newcommand{\light}{{\mathrm{li}}}
\newcommand{\gauge}{{\mathrm{g}}}

\DeclareMathOperator{\Tr}{Tr}

\newcommand{\insfigscaled}[3]{

\medskip
\noindent
\begin{minipage}{\linewidth}

\makebox[\linewidth]{\includegraphics[keepaspectratio=true,scale=#2]{figures/#1.pdf}}

\captionof{figure}{#3}

\label{fig:#1}
\end{minipage}
\medskip

}

\begin{document}

\title{BPS states in the Minahan-Nemeschansky $E_6$ theory}
\date{}
\author[1]{Lotte Hollands}
\author[2]{Andrew Neitzke}
\affil[1]{Department of Mathematics, Heriot-Watt University}
\affil[2]{Department of Mathematics, University of Texas at Austin}

\maketitle

{\abstract{We use the method of spectral networks to
compute BPS state degeneracies in the Minahan-Nemeschansky $E_6$ theory,
on its Coulomb branch, without turning on a mass deformation.
The BPS multiplicities come out in representations of the $E_6$
flavor symmetry.
For example, along the simplest ray in electromagnetic charge
space, we give the first $14$ numerical degeneracies,
and the first $7$ degeneracies as representations of $E_6$.
We find a complicated spectrum, exhibiting exponential growth
of multiplicities as a function of the electromagnetic charge.
There is one unexpected outcome: the spectrum is
consistent (in a nontrivial way) with the hypothesis of
\ti{spin purity}, that if a BPS state in this theory
has electromagnetic charge equal to $n$ times a primitive charge,
then it appears in a spin-$\frac{n}{2}$ multiplet.
}}

\setcounter{page}{1}


\section{Introduction}

\subsection{Setup}

The $E_6$ Minahan-Nemeschansky theory $T_3$
is a $4$-dimensional $\N=2$ superconformal field theory,
first discovered in \cite{Minahan:1996fg},
and studied at great length since then
(e.g. for a few highlights see \cite{Seiberg:1996bd,Minahan:1996cj,Aharony:2007dj,Gadde:2010te,Gadde:2015xta}.)
In \cite{Gaiotto:2009we}, it was
shown that $T_3$ can be constructed as a
theory of class $\cS[A_2]$ (see \cite{Gaiotto:2009hg} for
more on the definition of class $\cS$).
More precisely, $T_3$ is the theory of class
$\cS[A_2]$ associated to the Riemann surface
$C = \C\PP^1 \setminus \{z_1,z_2,z_3\}$
with full punctures at the points $z_\ell$.
In this paper, we will use this description of theory $T_3$
extensively; indeed, we take it as our \ti{definition}
of $T_3$.

The construction of theories of class $\cS[A_2]$
has in the ultraviolet a
flavor symmetry group $F_\ell \simeq SU(3)$
for each full puncture $z_\ell$. In our case
this gives a total flavor symmetry $F_\cS \simeq SU(3)^3$.
Remarkably, upon flowing to the infrared to reach the superconformal
field theory $T_3$, this symmetry is enhanced to a group
$F \simeq E_6$ (the compact simply connected form).\footnote{$E_6$ does not have a subgroup isomorphic to $SU(3)^3$,
but does have one isomorphic to $SU(3)^3 / \Z_3$, where the $\Z_3$ is
the diagonal subgroup. Thus, this symmetry enhancement requires that a certain
$\Z_3$ subgroup of $F_\cS$ acts trivially in the infrared.}

Theory $T_3$ has a 1-dimensional Coulomb branch, parameterized
by $u \in \C$.
When we move onto the Coulomb branch the scale invariance
and $U(1)_R$ symmetry are spontaneously broken. These broken
symmetries together act by $u \mapsto \lambda u$ for
$\lambda \in \C^\times$. Thus the physics is the
same for any $u \neq 0$.
In the infrared, it is given
by $\N=2$ supersymmetric $U(1)$ gauge theory.
As we will explain below, the electromagnetic
charge lattice $\Gamma_\gauge$ has three distinguished elements
$\gamma_{1,2,3}$, with $\gamma_1 + \gamma_2 + \gamma_3 = 0$.
After choosing an electromagnetic duality frame we could call
one of these three ``electric,'' and the other two magnetic or dyonic. There is no canonical such choice, though, and indeed
there is a $\Z_3$ symmetry which cyclically permutes the $\gamma_i$.

\subsection{Summary}

In this paper we compute
counts of 4d BPS particles of theory $T_3$ on its
Coulomb branch. As we have noted above, all points on the
Coulomb branch are physically equivalent, so computing the counts at one point is enough to determine them everywhere on the Coulomb branch.
(In particular, there are no walls of marginal stability
where BPS bound states can form or decay.)

More precisely, we compute \ti{indexed} counts,
the second helicity supertraces $\Omega(\gamma)$,
for various charges $\gamma \in \Gamma_\gauge$. Our
main tool is the technology of \ti{spectral networks}
introduced in \cite{Gaiotto2012}.
Here is a summary of the results:

\begin{enumerate}
\item We give an algorithm which efficiently
determines $\Omega(n \gamma_1)$, at least for $1 \le n \le 200$.
For example, we find
\begin{equation}
\Omega(6\gamma_1) = -114204.
\end{equation}
See \S\ref{sec:multiplicities-n0} for the results with $1 \le n \le 14$.
We also give $\Omega(n (\gamma_1 + 2 \gamma_2))$ and
$\Omega(n (\gamma_1 + 3 \gamma_2))$, both for $1 \le n \le 13$,
in \S\ref{sec:results-higher-pq}.

\item
We show that we have the
asymptotic exponential growth
\begin{equation}
	\abs{\Omega(n \gamma_1)} \sim c n^{-\frac52} (7 + 4\sqrt{3})^n
\end{equation}
for a constant $c$.
See \S\ref{sec:multiplicities-n0}.

\item
Since the theory $T_3$ has unbroken $F \simeq E_6$ flavor symmetry
on the Coulomb branch, the $\Omega(\gamma)$
can be ``upgraded'' from integers to characters $\bom(\gamma)$
of (virtual) representations of $F \simeq E_6$.
We compute
\begin{itemize}
  \item $\bom(n\gamma_1)$ with $1 \le n \le 7$,
  \item $\bom(n (\gamma_{1} + 2 \gamma_2))$
for $1 \le n \le 3$,
\item $\bom(\gamma_1 + 3 \gamma_2)$.
\end{itemize}
See \S\ref{sec:multiplicities-n0} and \S\ref{sec:results-higher-pq}
for the results.
For example, we find
\begin{equation} \label{eq:bps-sample}
\bom(4 \gamma_1) = -4 \times \overline{\mathbf{351}} - 8 \times \overline{\mathbf{27}}.
\end{equation}
Note that we get $E_6$ symmetry although
at intermediate stages of the computation we use a surface
defect of the theory, which leaves manifest only
the group $F_\cS \simeq SU(3)^3$.
This gives a check of our formalism.

\item We show that BPS states exist with \ti{all} primitive charges,
i.e. charges $\gamma = p \gamma_1 + q \gamma_2$ with $(p,q) = 1$.
See \S\ref{sec:arbitrary-pq}.

\item Our results are consistent with a surprising
hypothesis, which we call \ti{spin purity}:
BPS states carrying electromagnetic charges which are $n$ times
a primitive charge are always in multiplets of spin $\frac{n}{2}$.
(So BPS states with primitive charge are always in hypermultiplets,
BPS states with 2 times the primitive charge are in vector multiplets,
and so on.) For example, in \eqref{eq:bps-sample} above,
the multiplicities appearing are $-4$ and $-8$:
these are both positive integer
multiples of $-4$, which is the contribution from a spin-$2$
multiplet.
See \S\ref{sec:spin-purity} for more on this.

\end{enumerate}

Here are some open problems:

\begin{enumerate}
\item
It would be interesting to give a direct \ti{proof}
within our formalism that the $SU(3)^3$ symmetry
in the BPS indices will always be enhanced to $E_6$.

\item It would also be nice to prove that the BPS indices
are all consistent with the spin purity hypothesis;
this is true for all indices we computed, but
we did not compute $\Omega(\gamma)$ for \ti{all}
charges.

\item Our evidence for spin purity in theory $T_3$
is circumstantial, because we only compute the BPS indices,
which is not enough to determine the spin content uniquely.
The paper \cite{Galakhov2014} gives an
extension of spectral network technology which can be used
to compute
the spin content of the BPS spectrum. Applying the methods
of that paper could give a stronger check or refutation of the
spin purity hypothesis in theory $T_3$.

\item
One might wonder more generally whether spin purity is
true in every superconformal field theory.
(As one small piece of evidence we note that it
does hold in the $SU(2)$ theory with $4$ fundamental
flavors, though in a more trivial way.)
Perhaps this can
be proven using the same technology recently applied to the
no-exotics conjecture \cite{noexotics}.

\item The methods used in this paper
in principle determine the full spectrum
for all charges $\gamma \in \Gamma_\gauge$,
including its $E_6$ representation content.
In practice, our algorithm is
rather computationally expensive, so that we have only been able to
compute the degeneracies $\bom(\gamma)$ for a few charges.
We are hopeful that with more cleverness
it would be possible to obtain closed forms, or at least
to compute for higher charges.
This might reveal more hidden structure.

\item In Section 5.3 of \cite{Huang2013} some BPS degeneracies
are given for a \ti{five-dimensional} theory, which upon $S^1$
compactification should reduce to the theory $T_3$ considered
here. They appear to be
related to the $\Omega(n \gamma_1)$ we compute, but the precise
relation remains an open question,
as explained in \S\ref{sec:multiplicities-n0} below. It would
also be interesting to understand the five-dimensional meaning
of the other $\Omega(\gamma)$ we compute (perhaps in terms
of strings in five dimensions rather than particles.)

\item In \cite{Alim:2011kw},
it is proposed that in a mass-deformed version of theory $T_3$
one can compute BPS degeneracies using quiver quantum
mechanics. In particular, \cite{Alim:2011kw} argues that there
is a point of the Coulomb branch of the mass-deformed theory
where the full spectrum
consists of $24$ hypermultiplets. This is far simpler than
the spectrum we are finding in the massless theory.
Nevertheless, using
the Kontsevich-Soibelman wall-crossing formula,
one could try to start from these
$24$ hypermultiplets and derive the spectrum we find
here; this would be a very interesting check.
(We have remarked above that there are no walls in the
Coulomb branch of the \ti{massless} theory; however, there
are plenty of walls in the larger parameter space where we
include masses as well as Coulomb branch parameters.)

\item The variant of the spectral network technique which we use here
should be applicable to other superconformal field theories as well.
For example, it would be interesting to analyze
the Minahan-Nemeschansky theories with
global symmetry $E_7$ and $E_8$ \cite{Minahan:1996cj};
this would give additional data to support
or refute the spin purity conjecture.

\end{enumerate}

\subsection*{Acknowledgements}

We thank Clay Cordova, Thomas Dumitrescu, Min-xin Huang, Sheldon Katz,
Albrecht Klemm, Pietro Longhi, Tom Mainiero and Chan Park for helpful discussions
and comments on draft versions of this paper.
LH's work is supported by a Royal Society Dorothy Hodgkin fellowship.
AN's work is supported by National Science Foundation
award 1151693.

\section{Seiberg-Witten geometry}

In this section we summarize the Seiberg-Witten geometry
of theory $T_3$ and fix notation.

As for any theory of class $\cS[A_2]$ with full punctures,
theory $T_3$ has a Coulomb branch, parameterized by
meromorphic cubic differentials $(\phi_2,\phi_3)$,
such that at each puncture $\phi_2$ has at most
a first-order pole, and $\phi_3$ at most a second-order pole.
Using the $PSL(2,\C)$ symmetry of $\C\PP^1$ we can fix the punctures to
be at $(z_1, z_2, z_3) = (1, \omega, \omega^2)$ with
\begin{equation}
	\omega = \E^{2 \pi \I/3}.
\end{equation}
Then the only allowed $\phi_2$, $\phi_3$ are
\begin{equation} \label{eq:differentials-massless}
	\phi_2 = 0, \qquad \phi_3 = - \frac{u \, \de z^3}{(z^3-1)^2}.
\end{equation}
Thus we have a $1$-dimensional Coulomb branch, parameterized
by $u \in \C$.
As we have mentioned in the introduction, the physics is the
same for any $u \neq 0$; from now on we fix $u=1$.

The $U(1)$ gauge theory which appears
in the infrared of theory $T_3$ on the Coulomb branch
is naturally described in terms of the
Seiberg-Witten curve,
\begin{equation}
	\Sigma = \{ \lambda^3 + \phi_3 = 0 \} \subset T^* C,
\end{equation}
or more concretely, in coordinates $(x,z)$ on $T^*C$
where $\lambda = x \, \de z$,
\begin{equation} \label{eq:Sigma}
	\Sigma = \left\{ x^3 - \frac{1}{(z^3-1)^2} = 0 \right\}.
\end{equation}
$\Sigma$ is a curve of genus $1$, with $3$ punctures.
The projection $\pi: \Sigma \to C$, $\pi(x,z) = z$,
is a $3$-fold covering, which is unbranched.

Filling in the punctures on $\Sigma$ we
obtain a smooth compact genus $1$ curve $\overline\Sigma$.
$\overline\Sigma$ is a branched covering of $\C\PP^1$,
with the branch points at the punctures.
We represent this covering concretely by gluing together $3$
copies of the complex plane along branch cuts,
as in Figure \ref{fig:spectral-cover}.

\insfigscaled{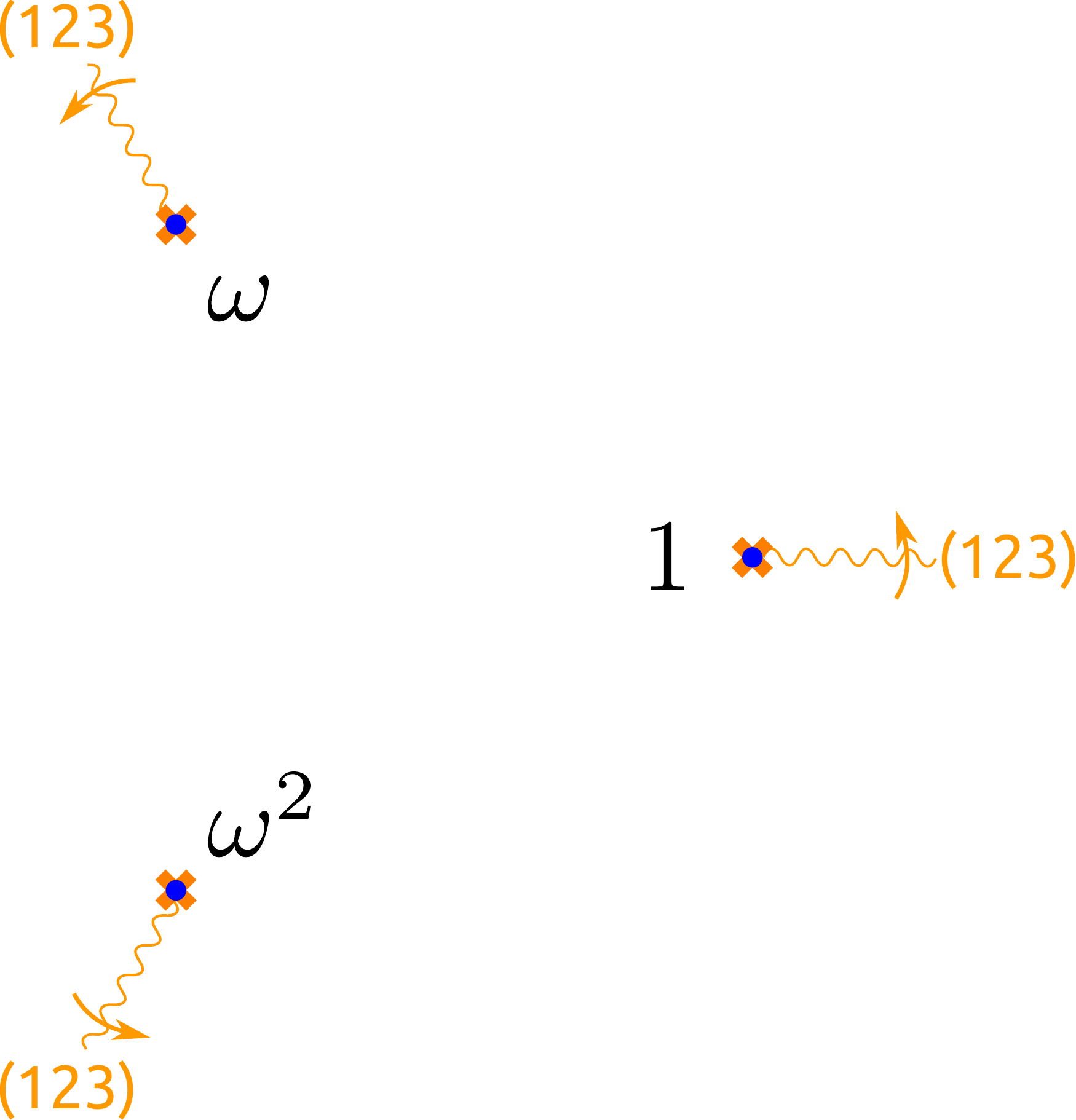}{0.22}{The thrice-punctured base curve
$C = \C\PP^1 \setminus \{1,\omega,\omega^2\}$,
in the inhomogeneous coordinate $z$.
The threefold cover $\Sigma \to C$ is obtained
by gluing together three sheets at branch cuts, represented
by wavy orange lines. The three branch
cuts meet at $z = \infty$. Each cut carries a sheet permutation $(123)$,
and a coorientation labeled by an arrow, which tells us which way to do the gluing:
sheet $1$ at the tail of the arrow is glued to sheet $2$ at the head of the
arrow, and so on.}

The $3$ sheets are labeled by the $3$ possible
choices of the cube root $\lambda$ of $-\phi_3$.
We fix the labeling as follows.
At $z = 0$ we have $-\phi_3 = \de z^3$.
We choose
\begin{equation}
 \lambda^{(1)}(z=0) = \omega \, \de z, \qquad  \lambda^{(2)}(z=0) = \omega^2 \, \de z, \qquad \lambda^{(3)}(z=0) = \de z.
\end{equation}

The electromagnetic charge lattice of the infrared gauge theory
on the Coulomb branch is
\begin{equation}
	\Gamma_\gauge = H_1(\overline\Sigma,\Z).
\end{equation}
As shown in Figure \ref{fig:spectral-cover-cycles} there are $3$ distinguished
charges $\gamma_{1,2,3} \in \Gamma_\gauge$, with
$\gamma_1 + \gamma_2 + \gamma_3 = 0$.

\insfigscaled{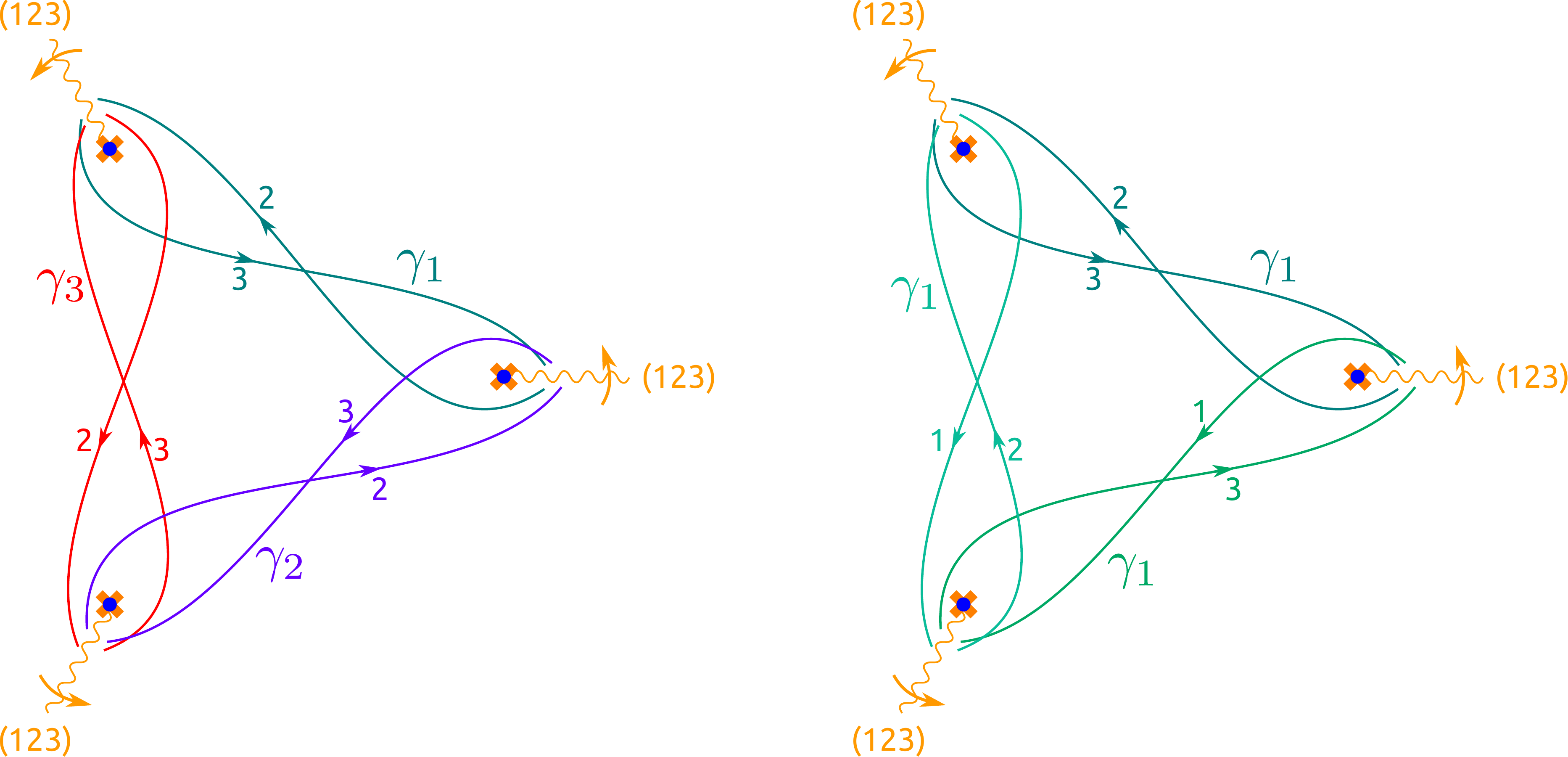}{0.33}{Cycles on $\overline\Sigma$. The
numbers next to path segments indicate which sheet of $\overline\Sigma$ the segments
lie on. Left: three cycles in homology classes $\gamma_1$, $\gamma_2$,
$\gamma_3$. Right: three cycles all in the same homology class
$\gamma_1$.}

The corresponding central charges $Z_\gamma = \frac{1}{\pi} \oint \lambda$ are
\begin{equation} \label{eq:periods}
	Z_{\gamma_1} = M, \qquad
	Z_{\gamma_2} = \omega^2 M, \qquad
	Z_{\gamma_3} = \omega M,
\end{equation}
where
\begin{align}
	M &= \frac{1}{\pi} \oint_{\gamma_1} \lambda
    = \frac{1}{\pi}\int_{1}^{\omega} \lambda^{(2)} + \frac{1}{\pi}\int_{\omega}^1 \lambda^{(3)} \\
    &= \frac{1}{\pi} (\omega^2 - 1) \int_{1}^{\omega} \lambda^{(3)} \\
&=  2^{-\frac23} \pi^{-\frac32} \Gamma\left(\frac13\right) \Gamma\left(\frac16\right) \approx 1.68702.
\end{align}

\section{Spectral networks}

Our computation of the spectrum uses the technology of
spectral networks, as described in \cite{Gaiotto2012},
slightly adjusted to deal with the case of
an unbroken nonabelian flavor symmetry.

For other applications of spectral networks to BPS
state counting, see
\cite{Gaiotto:2012db,Maruyoshi:2013fwa,Gabell2016,Longhi2016}.
Two useful tools for exploration of spectral networks
are the software package
{\tt {loom}} described in Section 5.1 of \cite{Longhi2016}
and the Mathematica notebook \cite{swn-plotter}. Both these
tools were used in developing the picture
described below.

\subsection{The canonical surface defect}

The key physical input to the definition of
spectral networks is the \ti{canonical surface defect}
\cite{Gaiotto:2009fs}.
This is a surface defect in theory $T_3$, which has $C$ as its
parameter space:
in other words we have a family of defects $\bbS_z$, $z \in C$.
We recall that $\bbS_z$ breaks the $\N=2, d=4$ supersymmetry
to $\N=(2,2), d=2$. Thus we can study
BPS particles living on $\bbS_z$ \cite{Gaiotto:2011tf},
whose properties are similar to those of BPS solitons in pure
$\N=(2,2), d=2$ field theories \cite{Cecotti:1992qh}, so we
also call them ``BPS solitons.''
In particular these BPS solitons carry a complex-valued central
charge $Z$.

The construction of $\bbS_z$ in
the class $\cS$ description of theory $T_3$
makes manifest that $\bbS_z$ preserves the $F_\cS$ flavor symmetry.\footnote{Indeed, $\bbS_z$ is reached by RG flow starting from a UV
description involving the 6d SCFT $\fX[A_2]$ with a $2$-dimensional
defect inserted at $z \in C$ and $4$-dimensional defects inserted
at the $z_\ell \in C$.
$F_\cS$ is already a symmetry of this UV description, living on the
$4$-dimensional defects.}
Thus we will find that the BPS solitons transform in representations
of $F_\cS \simeq SU(3)^3$. However, $\bbS_z$
does \ti{not} need to preserve the $F \simeq E_6$ symmetry which appears in the
infrared, and indeed we will find below that the BPS solitons on
$\bbS_z$ do not transform in representations of $F$.
Of course, the BPS states of
the 4d theory $T_3$ do transform in representations of $F$, as we
will verify in the examples we compute below.

\subsection{Spectral networks}

Now we recall the notion of spectral network.
There is one spectral network $\cW(\vartheta) \subset C$
for each phase $\vartheta$.
A point $z \in C$ lies on $\cW(\vartheta)$
if and only if $\bbS_z$ carries BPS
solitons of central charge $Z$, such that $-Z$ has phase $\vartheta$.
In this case we will say that $z \in C$ \ti{supports} these
BPS solitons. In general $\cW(\vartheta)$ is made up of curves
called \ti{walls}, each wall corresponding to a single
soliton charge.

In \cite{Gaiotto2012} an algorithm was described for
determining $\cW(\vartheta)$.
The key idea is first to restrict attention to solitons
which are lighter than some mass cutoff $\Lambda$, thus defining
a truncated network $\cW(\vartheta)[\Lambda]$.
For very small $\Lambda$ (much lighter than the masses of any BPS
particles in the $4$-dimensional theory),
$\cW(\vartheta)[\Lambda]$ is contained in the union of
small neighborhoods around the points where some solitons become massless.
There is a standard generic behavior around a point $z$
where only a single soliton becomes massless, which
we can use to determine $\cW(\vartheta)[\Lambda]$ for very small
$\Lambda$.
Then there is a scheme for determining how $\cW(\vartheta)[\Lambda]$
evolves as $\Lambda$ is continuously increased: each wall ends at a ``tip''
where the soliton mass reaches $\Lambda$, and the walls grow
from their tips, according to a differential equation expressing
the condition that $\E^{-\I \vartheta} Z$ remains real.
When walls cross, additional walls can be born from the
intersection points, corresponding to bound states
formed between existing solitons.
See \cite{Gaiotto2012} for the details.
Taking the limit $\Lambda \to \infty$ produces the
desired $\cW(\vartheta)$.

For the purpose of studying 4d BPS states of charge $\gamma$,
we need to study the spectral network corresponding to the phase
\begin{equation}
\vartheta_\gamma = \arg(-Z_\gamma).
\end{equation}
Note that $\vartheta_\gamma = \vartheta_{n \gamma}$ for any
$n > 0$, so the single network $\cW(\vartheta_\gamma)$ contains
information about 4d BPS states with all
charges $\gamma, 2 \gamma, 3 \gamma, \dots$.

In the simple examples computed in \cite{Gaiotto2012}, each
spectral network was only responsible for finitely many
4d BPS states. However, in \cite{Galakhov2013} it was
found that in $\N=2$ super Yang-Mills with gauge group $SU(3)$,
a single network can give rise to
infinitely many 4d BPS states (though still finitely many with
each fixed charge). We will see below that this also happens
in theory $T_3$.

\subsection{Spectral networks in theory \texorpdfstring{$T_3$}{T3}}

Because of the unbroken flavor symmetry in the 4-dimensional
theory $T_3$, the networks $\cW(\vartheta)$
have some special features. In particular, when $z$ approaches any
of the punctures $z_\ell$ on $C$, several solitons on $\bbS_z$
become massless at once.
In this situation, the rules of \cite{Gaiotto2012} cannot be applied
directly to determine $\cW(\vartheta)$.

Nevertheless,
we can determine the shape of $\cW(\vartheta)$
by first making a small deformation of
theory $T_3$ by a mass parameter $m$,
taking instead of \eqref{eq:differentials-massless}
\begin{equation} \label{eq:differentials-massive}
	\phi_2 = \frac{m z\, \de z^2}{(z^3 - 1)^2}, \qquad \phi_3 = - \frac{\de z^3}{(z^3-1)^2}.
\end{equation}
This deformation breaks the flavor symmetry to a Cartan
subgroup.
After making this perturbation, the
rules of \cite{Gaiotto2012} can be applied
to determine the network $\cW(\vartheta)$.
Then we can determine $\cW(\vartheta)$ in the massless
theory by taking the limit $m \to 0$. We will see
an example momentarily.

\subsection{The circle network}

To make our discussion more concrete,
we now specialize to a specific phase, $\vartheta = \pi$.
This will lead to a particularly simple spectral
network $\cW(\vartheta)$.
Since $\vartheta_{\gamma_1} = \pi$ this is the phase relevant
for studying BPS states of charge $n \gamma_1$, $n > 0$.
In \S\ref{sec:other-charges} below, we will
consider more general charges and networks.

We begin by studying $\cW(\vartheta)$ for $\vartheta$
slightly perturbed from $\pi$.
Figure \ref{fig:deformed-circle-network}
shows a sample network $\cW(\vartheta)[\Lambda]$, obtained
with the help of a computer and the Mathematica notebook \cite{swn-plotter}.

\insfigscaled{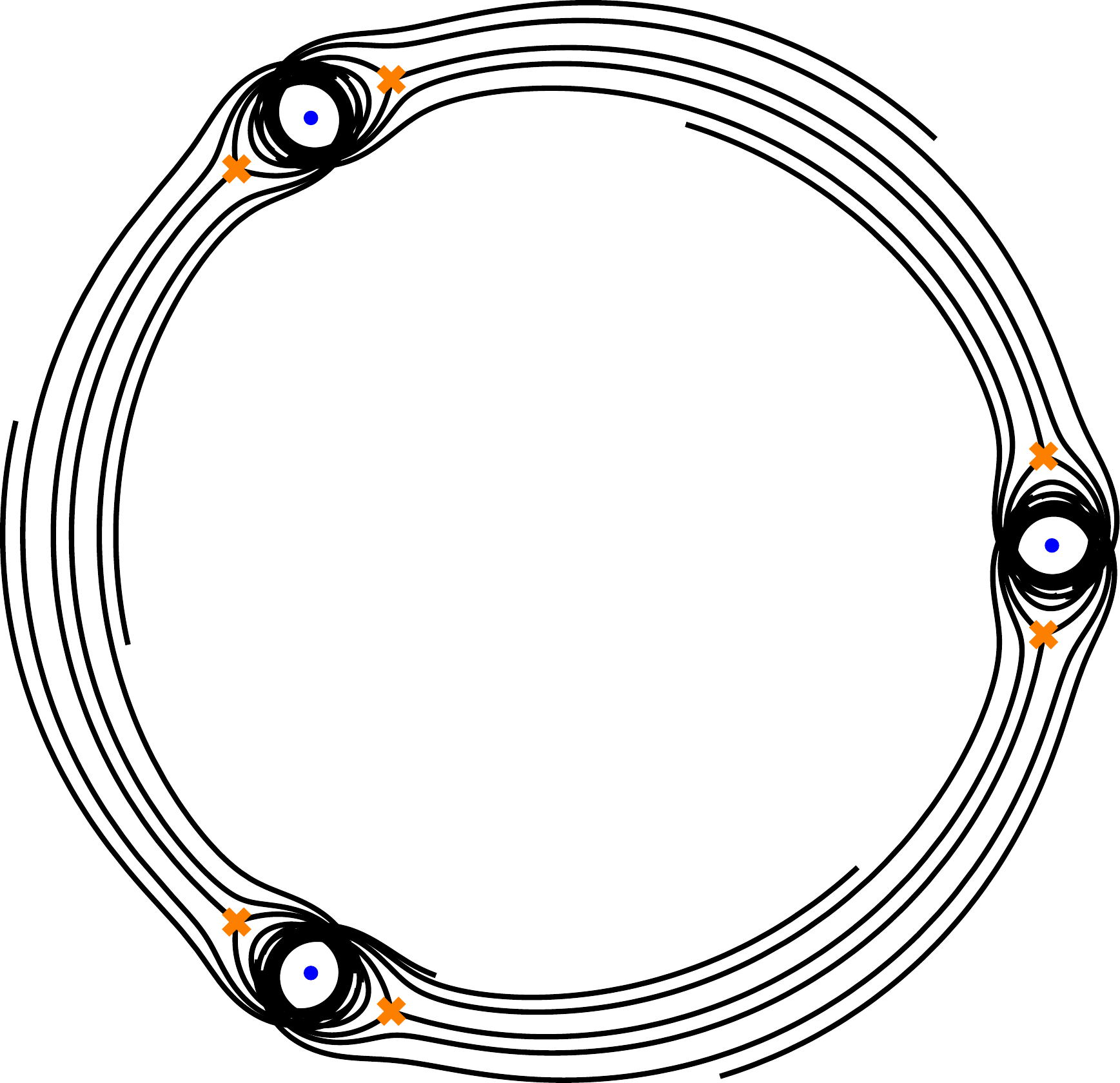}{0.25}{The spectral network
$\cW(\vartheta = \pi+0.02)[\Lambda = 1.5]$, where we took $m = 1.25$
in \eqref{eq:differentials-massive}.
The blue dots are the punctures, and the orange crosses are
branch points of the deformed spectral cover $\Sigma$.}

Looking at Figure
\ref{fig:deformed-circle-network},
we notice immediately that it is ``almost degenerate,'' in the sense
that there are groups of walls which are close together. The reason
for this is that there are various solitons
whose electromagnetic charges differ by some
multiple of $\gamma_1$; their central charges
thus differ by a real number (in fact an integer multiple of $M$).
In the limit where we take $\vartheta \to \pi$,
the walls supporting these solitons merge into a single wall,
which now supports infinitely many distinct solitons, with masses
diverging to $\infty$.
In Figure \ref{fig:deformed-circle-network-2} we show the behavior
very near $\vartheta = \pi$.

\insfigscaled{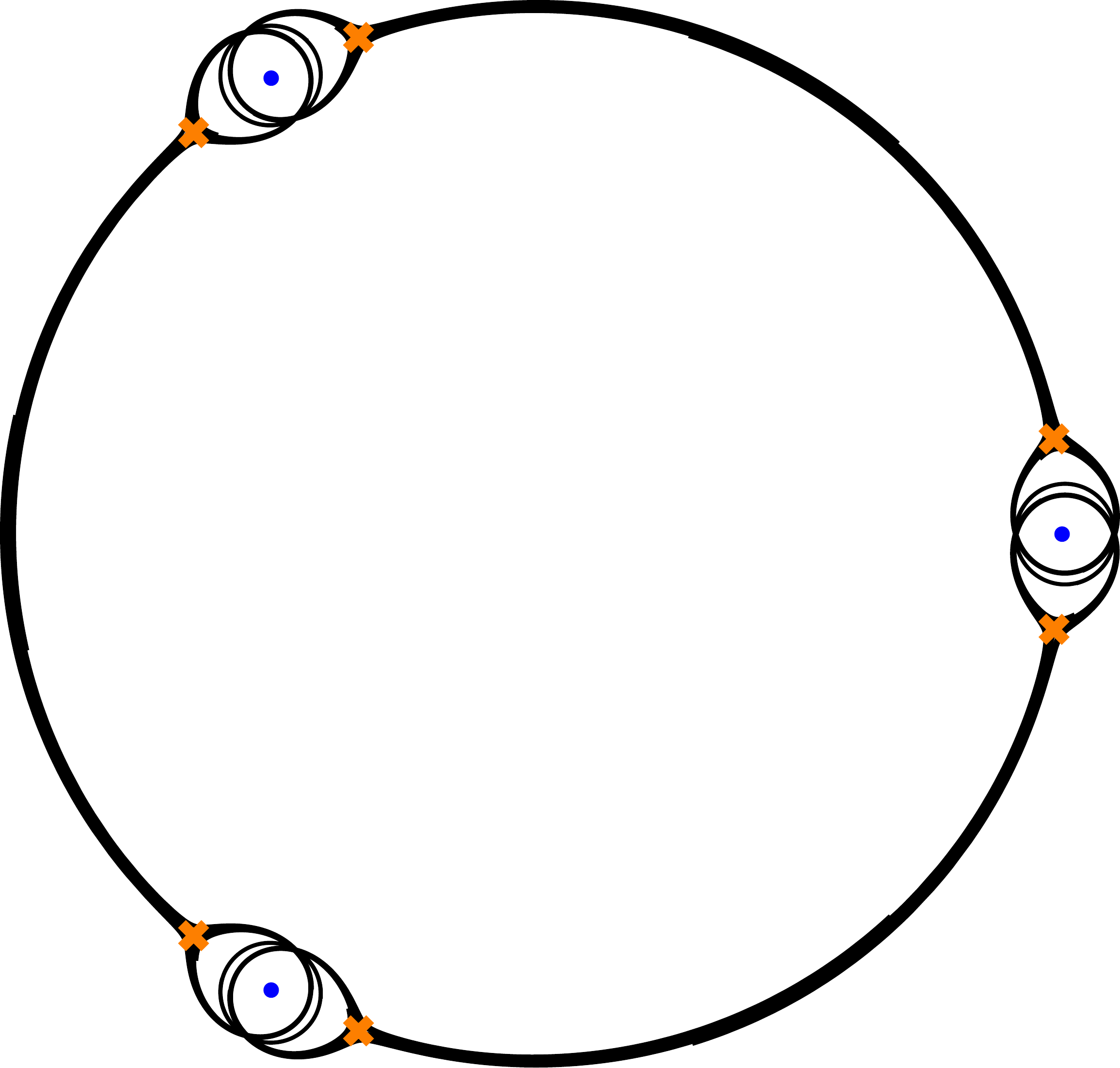}{0.24}{The spectral network $\cW(\vartheta = \pi+0.002)[\Lambda = 1.5]$, where we took $m = 1.25$ in \eqref{eq:differentials-massive}. The blue dots are the punctures, and the orange crosses are branch points of the deformed spectral cover $\Sigma$.}

Then taking the mass deformation $m \to 0$, the figure is further
simplified, because the branch points move onto the punctures.
Thus we arrive at the
network of Figure \ref{fig:circle-network}.
It consists of three walls connecting the
three punctures. The three walls together make
up the equator of $C$.

\insfigscaled{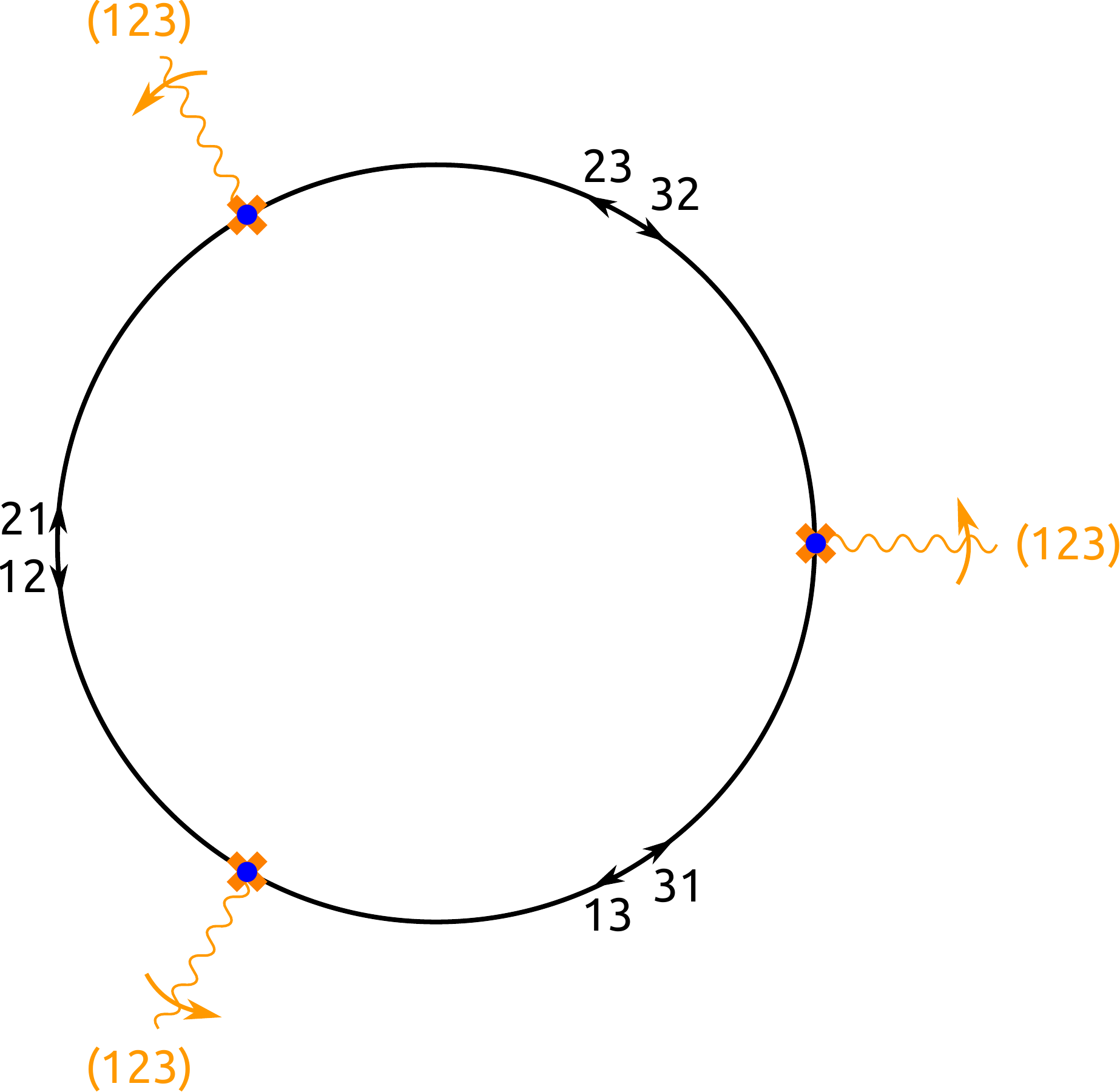}{0.4}{The spectral network $\cW(\vartheta = \pi)$.}

The labels $ij$ on each wall in Figure \ref{fig:circle-network} tell us
which types of BPS solitons can occur there. We call a soliton
``of type $ij$'' if it interpolates between the vacuum of $\bbS_z$ labeled
$i$ (at $x \to -\infty$) and the vacuum labeled $j$ (at $x \to +\infty$).
If $z$ is on
a wall of $\cW(\vartheta = \pi)$
carrying the label $ij$, then there exists a BPS soliton of
type $ij$ on $\bbS_z$
with central charge $Z \in \R_+$.
As we move $z$ along a wall of $\cW(\vartheta = \pi)$, the central charge
of these BPS solitons changes, while remaining real and positive.
The arrow next to a label $ij$ in Figure \ref{fig:circle-network}
indicates the direction in which the central charge is increasing,
for solitons of type $ij$.

As we see from Figure \ref{fig:circle-network},
each of the three walls in $\cW(\vartheta = \pi)$ is supporting
solitons of two complementary types, $ij$ and $ji$. We call walls which
support two complementary types of solitons
``double walls'' (previously
``two-way streets'' in the parlance of \cite{Gaiotto2012}.)

\section{Computing the soliton spectrum} \label{sec:computing-solitons}

\subsection{Counting BPS solitons} \label{sec:counting-solitons}

We now consider the spectrum
of BPS solitons supported on the defect $\bbS_z$, when
$z$ lies on $\cW(\vartheta = \pi)$.
Actually, if we take precisely $\vartheta = \pi$
this spectrum is ill-defined, due to mixing with BPS states
of the bulk 4-dimensional theory carrying charge $n \gamma_1$.
This mixing is precisely
what we want to study, as an indirect way of determining
the spectrum of bulk BPS states. So what we do
(again following the rules of \cite{Gaiotto2012})
is to study not $\vartheta = \pi$ on the nose
but rather the two limits $\vartheta \to \pi^\pm$.
These correspond to two ``resolutions'' of the network,
as shown in Figure \ref{fig:circle-network-resolved}.
In each of the two resolutions, each double wall is split into
two infinitesimally separated walls.

\insfigscaled{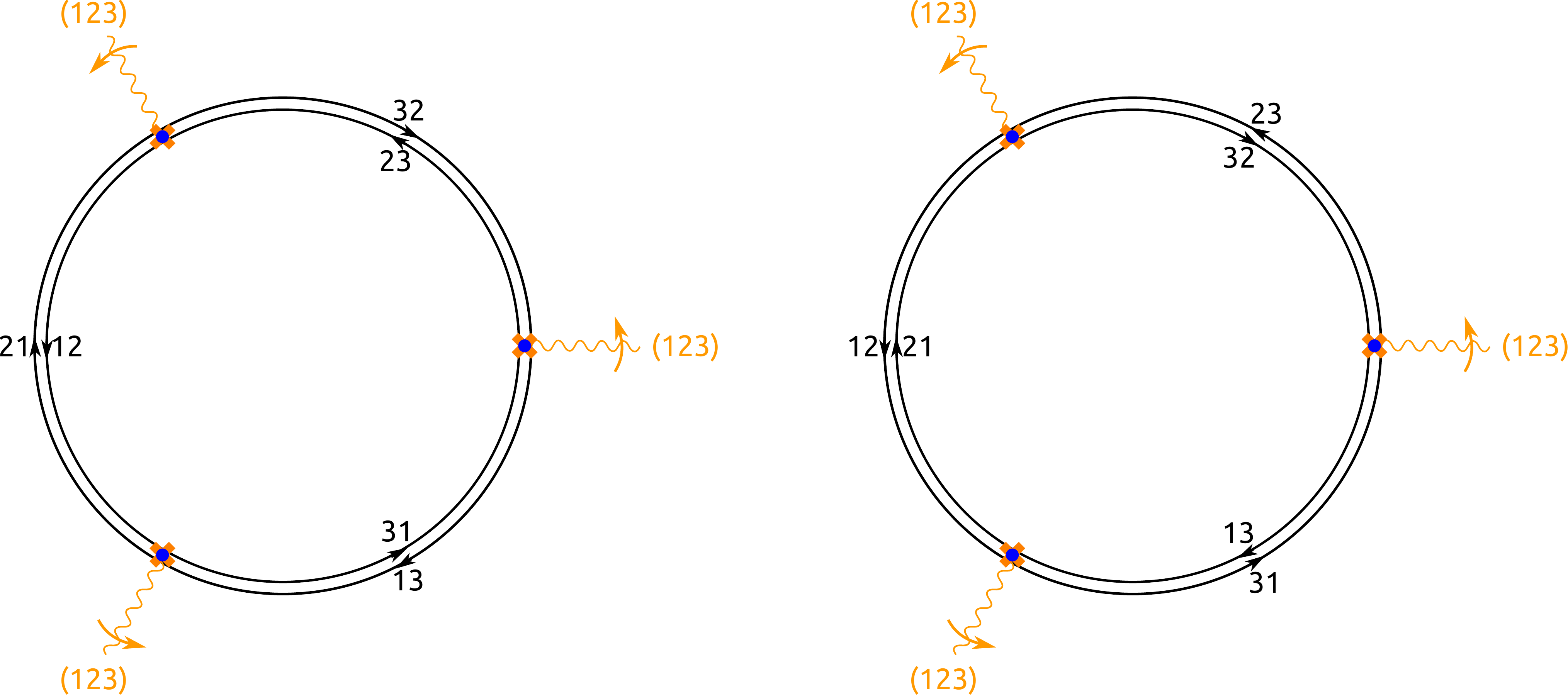}{0.3}{The two resolutions
of the spectral network $\cW(\vartheta = \pi)$.}

As we have explained,
in general there can be multiple solitons with $Z \in \R_+$
on the same $\bbS_z$.
The Hilbert space $\cH_z$ of BPS solitons on the defect $\bbS_z$
is decomposed into sectors $\cH_{z,ij}$, where $i,j$ label the
asymptotic vacua at spatial infinity. In addition, different
solitons in $\cH_{z,ij}$ can carry different electromagnetic
charges, which we label $a$.
Thus $\cH_{z,ij}$ is decomposed into charge sectors $\cH_{z,ij,a}$.

The electromagnetic charges $a$ of the solitons
are not integrally quantized: rather, they have a
fractional part,
determined by the parameter $z$ of the surface defect
as well as the vacua $i$, $j$ connected by the soliton,
as explained in \cite{Gaiotto:2011tf}.
Still, we may have different solitons with the same values of $i,j,z$
and in this case their electromagnetic charges do differ by a
quantized charge as usual.
The invariant way of describing this is to say that the charges
$a \in \Gamma_{ij,zz}$ where $\Gamma_{ij,zz}$ is a \ti{torsor}
for the lattice $\Gamma_\gauge$. Concretely, $\Gamma_{ij,zz}$ can be
identified with the set of homology classes of \ti{open} paths
on $\overline\Sigma$, running from $z^{(i)}$ to $z^{(j)}$.
The central charge of a soliton with charge $a$ is
\begin{equation}
	Z = Z_a = \frac{1}{\pi} \int_a \lambda
\end{equation}
where $\lambda \in \Omega^{1,0}(\Sigma)$ is the tautological
1-form.

We count the BPS solitons by a supersymmetric index
depending on a flavor parameter $g \in F_\cS \simeq SU(3)^3$:
\begin{equation}
  \mu_{z,ij,a}(g) = \Tr_{\cH_{z,ij,a}} F(-1)^F g.
\end{equation}
When $g=1$ this reduces to the index considered in \cite{Cecotti:1992qh};
short multiplets of $\N=(2,2)$ supersymmetry
($2$ states) contribute $\pm 1$ to this index
depending on whether the ground state has fermion number
$F$ even or odd; long multiplets ($4$ states) contribute $0$.
For general $g$, we use the decomposition of $\cH_{z,ij,a}$ as
a direct sum of multiplets $V \otimes W$, where
$V$ is a representation of
$F_\cS$ and $W$ a representation of $\N=(2,2)$
supersymmetry; if $W$ is short the multiplet $V \otimes W$
contributes $\pm \Tr_V g$, and if $W$ is long
it contributes $0$.

It is convenient to package the spectrum
into a formal generating function:
\begin{equation}
  \bS_z = \sum_{a \in \Gamma_{ij,zz}} \mu_{z,ij,a} X_a,
\end{equation}
where $X_{a}$ is a formal variable, obeying the multiplicative relation
$X_b X_a = X_{a+b}$ if $a$ and $b$ are paths which can be concatenated
as in Figure \ref{fig:concatenation},
and $X_a X_b = 0$ otherwise.\footnote{More succinctly, these
formal variables live in the \ti{groupoid ring} corresponding to
the groupoid of open paths on $\Sigma$.}

\insfigscaled{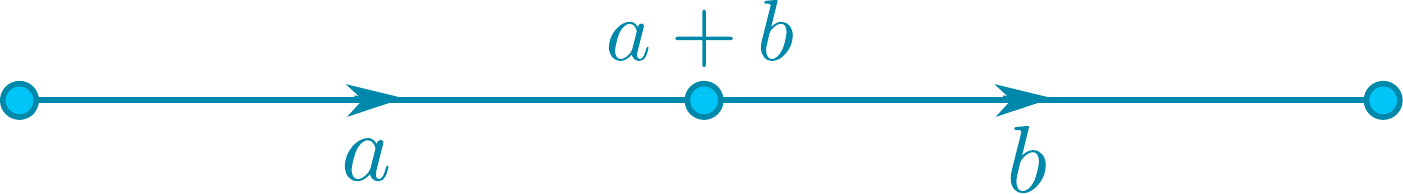}{0.43}{Two paths $a$, $b$ on $\Sigma$ which can
be concatenated; the concatenated path is $a+b$.}

The generating functions $\bS_z$ for different points $z$, $z'$ on the same
wall are related to one another by ``continuation'' or ``Gauss-Manin
connection'': as we continuously deform $z$ to $z'$, the paths
$a \in \Gamma_{ij,zz}$ continuously deform into paths $a' \in \Gamma_{ij,z'z'}$, and the degeneracies $\mu_{z,ij,a}$ remain constant.
More formally, we can describe this as follows.
Let $I$ denote the wall segment running from $z$ to $z'$,
and let $I^{-1}$ be the reverse of $I$.
Let $\bF_I$ denote the sum of the $3$ lifts of $I$ to $\Sigma$,
\begin{equation}
 \bF_I = \sum_{i=1}^3 X_{I^{(i)}}.
\end{equation}
and similarly $\bF_{I^{-1}}$. Then
\begin{equation}
  \bS_{z'} = \bF_{I} \bS_z \bF_{I^{-1}} .
\end{equation}

\subsection{Framed 2d-4d BPS states}

The constraints which we use to determine the
$\bS_z$ arise from consideration of
yet another kind of BPS states, the \ti{framed 2d-4d BPS states}
attached to supersymmetric \ti{interfaces} between
surface defects $\bbS_z$ and $\bbS_{z'}$ \cite{Gaiotto:2011tf}.

Given a path $\wp$ on $C$ between $z$ and $z'$ and a phase $\vartheta$,
there is a corresponding supersymmetric interface $\bbL_{\wp,\vartheta}$.
Since we are fixing $\vartheta = \pi$ until
\S\ref{sec:other-charges}, we abbreviate this as $\bbL_\wp$.
The framed 2d-4d BPS states on $\bbL_\wp$ make up a Hilbert space
$\cH_\wp$, which decomposes similarly to the Hilbert space
$\cH_z$ we considered above. First,
$\cH_\wp$ decomposes into sectors $\cH_{\wp,ij}$
labeled by the asymptotic vacua $ij$.
Second, $\cH_{\wp,ij}$ decomposes into sectors $\cH_{\wp,ij,a}$
labeled by the possible 4d electromagnetic charges
$a \in \Gamma_{ij,zz'}$ where $\Gamma_{ij,zz'}$ is
the space of homology classes of \ti{open} paths
on $\overline\Sigma$, running from $z^{(i)}$ to $z'^{(j)}$.
Finally, $\cH_{\wp,ij,a}$ is a representation of the
flavor symmetry $F_\cS \simeq SU(3)^3$.
We count the framed 2d-4d BPS states by a supersymmetric index
which is a function on
$F_\cS$:
\begin{equation}
  \fro_{\wp,ij,a}(g) = \Tr_{\cH_{\wp,ij,a}} (-1)^F g.
\end{equation}
It is convenient to package the spectrum into a generating function
\begin{equation}
  \bF_\wp = \sum_{i,j=1}^3 \bF_{\wp,ij}, \qquad \bF_{\wp,ij} = \sum_{a \in \Gamma_{ij,zz'}} \fro_{\wp,ij,a} X_{a}
\end{equation}
where the $X_a$ are formal variables as above.

\subsection{Constraints on framed 2d-4d BPS states} \label{sec:2d4d-constraints}

The spectrum of framed 2d-4d BPS states on interfaces $\bbL_\wp$ obeys
constraints which are easily summarized in terms of the corresponding
$\bF_\wp$:
\begin{enumerate}
  \item If $\wp$ and $\wp'$ are two paths which are homotopic,
  then
  \begin{equation}
    \bF_\wp = \bF_{\wp'}.
  \end{equation}

  \item If $\wp$ and $\wp'$ are two paths which can be
  concatenated, so that the end of $\wp'$ equals the start
  of $\wp$, then
  \begin{equation}
    \bF_{\wp \wp'} = \bF_\wp \bF_{\wp'}.
  \end{equation}

  \item If $\wp$ is a path which does not cross any of the
  walls, then
  \begin{equation}
    \bF_\wp = \sum_{i=1}^3 X_{\wp^{(i)}}
  \end{equation}
  where $\wp^{(i)}$ denote the $3$ lifts of the path $\wp$
  to $\Sigma$.

  \item If $\wp$ is a path which crosses just a single wall,
 then
  \begin{equation} \label{eq:path-crossing}
    \bF_\wp = \bF_{\wp_-} (1 \pm \bS_z) \bF_{\wp_+},
  \end{equation}
  where $\wp_\pm$ are the two segments of $\wp$
  as shown in Figure \ref{fig:path-crossing-wall}, and $z$ is the intersection point.

 \insfigscaled{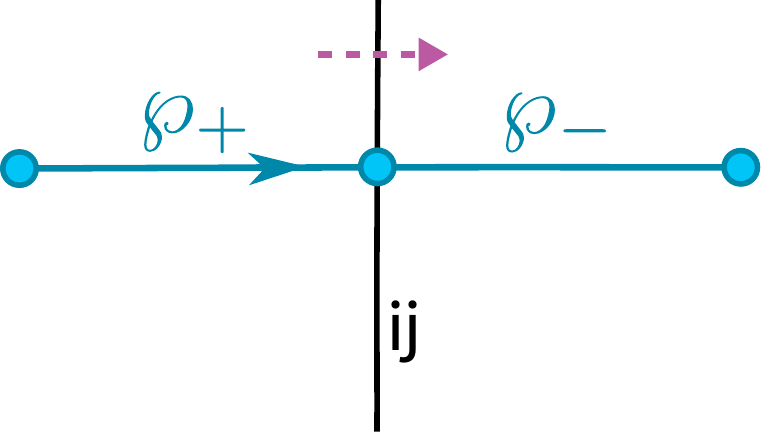}{0.45}{The path $\wp$ crossing
 a wall, divided into two segments $\wp_\pm$. The dotted arrow
 indicates a choice of coorientation of the wall.}

  The sign $\pm$ in \eqref{eq:path-crossing} is controlled by
  a subtle point which we have suppressed until now: in general there are
  ambiguities in defining the fermion number operator $F$ which
  appears in the definition of $\mu$.
  In \cite{Gaiotto2012} a scheme for fixing these ambiguities was proposed,
  and we assume here that this scheme is correct; this means that
  to fix the ambiguity we need to choose a coorientation
  of each wall in $\cW$. The
  sign in \eqref{eq:path-crossing} is $+$ if we cross in the direction
  given by the coorientation, and $-$ if we cross in the opposite direction.
	In Figure \ref{fig:path-crossing-wall} the dotted arrow indicates
	one possible choice of coorientation; for this choice the sign
	in \eqref{eq:path-crossing} would be $+$.

  \item \label{item:puncture-constraint}
  Let $\wp$ be a loop which goes counterclockwise
  around puncture $z_\ell$. Let $\bM_\wp(g)$
  denote the matrix $\{\bF_{\wp,ij}(g)\}_{i,j=1}^3$.
  Then $\Tr \bM_\wp(g)$ is a linear combination of the formal variables
  $X_a$ for $a \in \Gamma_{ii,zz}$. Passing from
  $a \in \Gamma_{ii,zz}$ to the corresponding class
  $\gamma \in \Gamma_\gauge$   (``forgetting'' the basepoint $z^{(i)}$)
  we replace these formal variables by formal variables $X_\gamma$, $\gamma \in \Gamma_\gauge$,
  now lying in a commutative algebra, with the simple relation
  $X_{\gamma} X_{\gamma'} = X_{\gamma + \gamma'}$. In particular
  $X_{\gamma = 0}$ behaves as the identity, so we write
  $X_{\gamma = 0} = 1$.

  Now we can formulate our condition around the puncture.
  First, it says that only the trivial element
  $X_{\gamma = 0}$ occurs in $\Tr \bM_\wp(g)$.
  Since $X_{\gamma=0} = 1$ this means
  we can interpret $\Tr \bM_\wp(g)$ simply as a number.
  Second, the condition says that this number is fixed, as follows.
  Let $R_\ell$ denote a $3$-dimensional representation of the
  flavor symmetry group $F_\cS = F_1 \times F_2 \times F_3$, in which $F_\ell \simeq SU(3)$
  acts via the fundamental representation ${\mathbf 3}$,
  and the other two $F_{\ell'}$ act trivially.
  Then:
  \footnote{To motivate this condition,
  note that from the closed loop $\wp$ we could construct a bulk line defect
  $L_\wp$ in theory $T_3$, not living on any surface defect
  \cite{Drukker:2010jp,Drukker:2009tz,Gaiotto:2010be}. This line defect
  is a flavor Wilson line in representation $R_\ell$. Likewise the
  loop $\wp^{-1}$ is a flavor Wilson line in representation $\overline{R}_\ell$.
  Upon gluing the
  puncture $z_\ell$ to another puncture, thus gauging the
  subgroup $F_\ell$, these would become honest gauge Wilson lines.
  There is a notion of framed BPS states for supersymmetric line defects
  \cite{Gaiotto:2010be}, and for a flavor Wilson line in any representation $R$,
  the space of framed BPS states is a copy
  of $R$, with zero electromagnetic charge; this leads to the conditions \eqref{eq:trace-condition}.}
  \begin{equation} \label{eq:trace-condition}
    \Tr \bM_\wp(g) = \Tr_{R_\ell} g, \qquad \Tr \bM_\wp(g)^{-1} = \Tr_{\overline{R}_\ell} g.
  \end{equation}
  These two conditions together are
  equivalent to requiring that the characteristic polynomial of
  $\bM_\wp(g)$ equals the characteristic polynomial of $g$ acting in representation $R_\ell$.
\end{enumerate}

\subsection{The constraint equations}\label{sec:constrainteqns}

Perhaps surprisingly, the constraints of \S\ref{sec:2d4d-constraints}
are strong enough to determine all of the soliton generating functions
$\bS_z$.
To see how this works, we first consider the local picture we obtain
by zooming in around one of the punctures on the left side of Figure
\ref{fig:circle-network-resolved}. This picture is indicated in Figure
\ref{fig:local-puncture-resolved}.

\insfigscaled{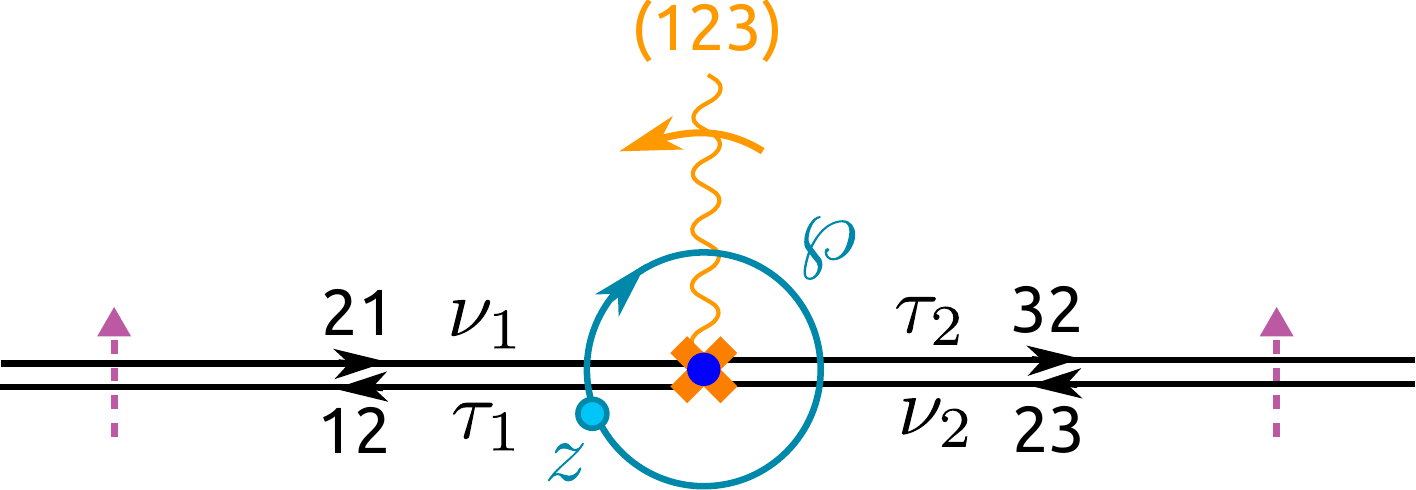}{0.45}{The local picture around
one of the punctures on the left side of Figure \ref{fig:circle-network-resolved}.}

More precisely, this is the picture for the puncture at $z = \omega$;
for the other punctures we would act by a cyclic permutation on the
sheet labels $123$. This permutation does not affect any of the following
computations.

We have labeled the walls in Figure \ref{fig:local-puncture-resolved}
with the symbols $\tau_1, \tau_2, \nu_1, \nu_2$ which we use to
represent the $\bS_z$ on these four walls.
In Figure \ref{fig:local-puncture-resolved} we have also
marked a loop $\wp$ on $C$, beginning and ending at the
marked point $z$. According to the constraints of \S\ref{sec:2d4d-constraints}, we have
\begin{equation}
	\bF_\wp = \bF_{\wp_-} (1 - \nu_2) (1 - \tau_2) \bF_{\wp_+} (1 + \nu_1) (1 + \tau_1).
\end{equation}
Here the $\nu$'s and $\tau$'s are evaluated at the
places where $\wp$ crosses the walls.
$\bF_{\wp_+}$ is the sum of three terms for the three
lifts of $\wp_+$ to $\Sigma$, and likewise $\bF_{\wp_-}$:
\begin{equation}
	\bF_{\wp_-} = x_{11} + x_{22} + x_{33}, \qquad \bF_{\wp_+} = x_{13} + x_{21} + x_{32},
\end{equation}
where each $x_{ij} = X_a$ for $a$ a path beginning on sheet $i$
and ending on sheet $j$.

Thus we can write
\begin{multline} \label{eq:Mp-long}
\bM_\wp = \\
	\begin{pmatrix}
	x_{11} & 0 & 0 \\
	0 & x_{22} & 0 \\
	0 & 0 & x_{33}
	\end{pmatrix}
  \begin{pmatrix}
    1 & 0 & 0 \\
    0 & 1 & 0 \\
    0 & -\nu_2 & 1
  \end{pmatrix}
  \begin{pmatrix}
    1 & 0 & 0 \\
    0 & 1 & -\tau_2 \\
    0 & 0 & 1
  \end{pmatrix}
  \begin{pmatrix}
    0 & x_{21} & 0 \\
    0 & 0 & x_{32} \\
    x_{13} & 0 & 0
  \end{pmatrix}
  \begin{pmatrix}
    1 & \nu_1 & 0 \\
    0 & 1 & 0 \\
    0 & 0 & 1
  \end{pmatrix}
  \begin{pmatrix}
    1 & 0 & 0 \\
    \tau_1 & 1 & 0 \\
    0 & 0 & 1
  \end{pmatrix}
  \\
  = \begin{pmatrix}
    x_{11} x_{21} \tau_1 & x_{11}x_{21} & 0 \\
    - x_{22} \tau_2 x_{13}(1 + \nu_1 \tau_1)  & - x_{22} \tau_2 x_{13} \nu_1  & x_{22} x_{32} \\
    x_{33} (1+\nu_2 \tau_2) x_{13} (1+\nu_1 \tau_1) & x_{33} (1 + \nu_2 \tau_2) x_{13} \nu_1 & - x_{33} \nu_2 x_{32}
  \end{pmatrix}.
\end{multline}
This formula is hard to read because of the proliferation of
$x_{ij}$'s. Fortunately,
we can safely set all
$x_{ij} = 1$, at the cost of remembering that $x_{ij}$
may need to be inserted to make
the products well defined and nonzero;
there is always a unique way of doing so.
From now on we adopt this convention. Then we can replace
\eqref{eq:Mp-long} by the simpler
\begin{equation}
	\bM_\wp
	 = \begin{pmatrix}
    \tau_1 & 1 & 0 \\
    -\tau_2(1 + \nu_1 \tau_1)  & - \tau_2 \nu_1  & 1 \\
    (1+\nu_2 \tau_2)(1+\nu_1 \tau_1) & (1 + \nu_2 \tau_2) \nu_1 & -\nu_2
  \end{pmatrix}.
\end{equation}
This gives for the characteristic polynomial
\begin{equation} \label{eq:charpoly-1}
  P(t) = 1 + t (\nu_1 - \tau_2 + \nu_2 \tau_1) + t^2 (-\nu_2 - \tau_2 \nu_1 + \tau_1) - t^3.
\end{equation}
According to constraint \ref{item:puncture-constraint} of \S\ref{sec:2d4d-constraints},
this is supposed to be equal to
\begin{equation} \label{eq:charpoly-2}
  P(t) = \prod_{k=1}^3 (\xi_k^{-1} - t) = 1 - A t + B t^2 - t^3,
\end{equation}
where $\xi_k \in \C^\times$ are the $3$ eigenvalues of $g$ in representation
$R_\ell$, satisfying $\prod_{k=1}^3 \xi_k = 1$, and we defined
$A = \sum_{k=1}^3 \xi_k$, $B = \sum_{k=1}^3 \xi_k^{-1}$.
Comparing \eqref{eq:charpoly-1} and \eqref{eq:charpoly-2}
determines the ``outgoing'' $\tau_n$ in terms of the ``incoming''
$\nu_n$:\footnote{To be pedantic, in writing \eqref{eq:charpoly-1} we have implicitly
inserted the appropriate $x_{ij}$'s to make $\tau_n$ and $\nu_n$ closed, and then
applied the closure operation to pass from the formal variables $X_a$ to the $X_\gamma$,
as explained in the discussion of constraint \ref{item:puncture-constraint}. After so
doing, \eqref{eq:charpoly-1} and
\eqref{eq:charpoly-2} are both equations in the commutative algebra
generated by the $X_\gamma$, and we can compute in the usual way.
This gives a version of \eqref{eq:puncture-traffic-rule} with the closure operation
applied to all variables. From this version we can uniquely recover
\eqref{eq:puncture-traffic-rule}.}
\begin{equation} \label{eq:puncture-traffic-rule}
\tau_1 = \frac{B + A \nu_1 + \nu_1^2 + \nu_2}{1 - \nu_1 \nu_2}, \qquad \tau_2 = \frac{A + B \nu_2 + \nu_2^2 + \nu_1}{1 - \nu_1 \nu_2},
\end{equation}
where the denominator is to be expanded in a geometric series.
These equations are the key to computing all the soliton counts, as we will see momentarily.

\subsection{A decoupled limit} \label{sec:decoupled-limit}

It is interesting to consider what
\eqref{eq:puncture-traffic-rule} would imply if
we assume that both $\nu_n = 0$. In this case we just get
\begin{equation} \label{eq:tau-isolated}
  \tau_1 = B, \qquad \tau_2 = A.
\end{equation}
In other words, each of the two outgoing walls supports $3$ solitons,
transforming in either the ${\bf 3}$ or $\overline {\bf 3}$ of the
flavor symmetry $F_\ell \simeq SU(3)$.
We might also write this as
\begin{equation} \label{eq:tau-isolated-2}
  \tau_1 = \overline{\bf 3}, \qquad \tau_2 = {\bf 3}.
\end{equation}
Thus the situation is as in Figure \ref{fig:decoupled-puncture}.

\insfigscaled{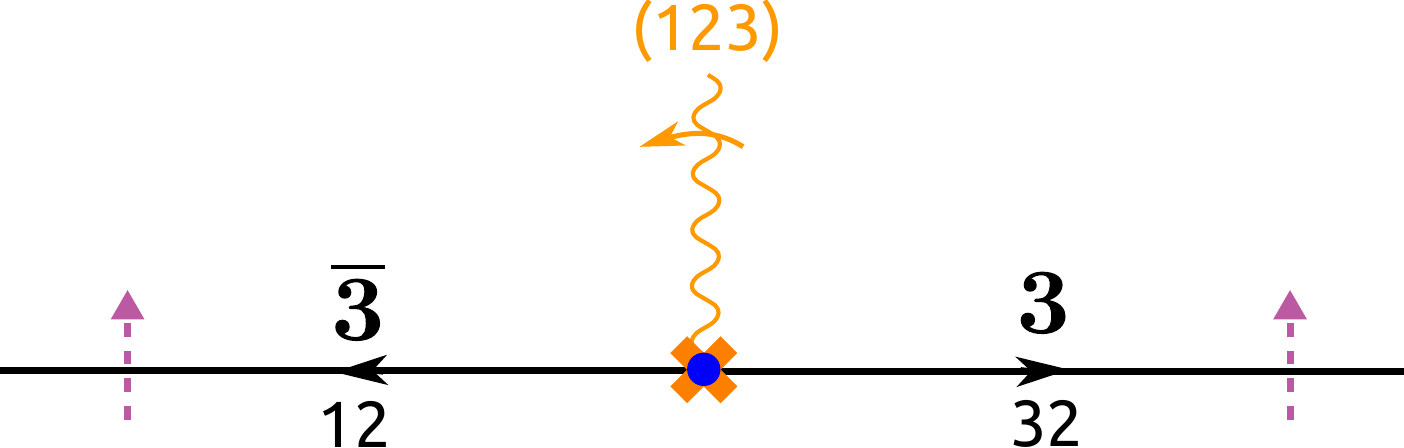}{0.5}{The spectral
network and soliton counts in a neighborhood of a puncture $z_\ell$,
including their transformation under the flavor symmetry
$F_\ell \simeq SU(3)$, if we assume that
there are no walls incoming.}

This result has a direct physical interpretation.
In the limit $z \to z_\ell$,
the central charges for some of the
BPS solitons on the surface defect $\bbS_z$ go to zero.
If we concentrate just on this light sector,
$\bbS_z$ appears to be a 2d theory decoupled from the 4d bulk, namely the
supersymmetric sigma model into $\C\PP^2$. Indeed, the soliton
spectrum of this model precisely consists of the ${\bf 3} \oplus {\overline{\bf 3}}$
of the $SU(3)$ flavor symmetry \cite{Koberle:1980wr}.

We can also see these $3 + 3$ solitons directly as follows.
We make a mass perturbation of the theory and
zoom in on a small neighborhood of one of the punctures.
The mass deformation splits the puncture into a puncture and two branch
points. A local model for the situation is obtained by taking the puncture
at $z=0$ and
\begin{equation} \label{eq:mass-perturbed-differentials}
  \phi_2 = \frac{3 m^2}{z^2} \de z^2, \qquad \phi_3 = \frac{2 \I}{z^2} \de z^3.
\end{equation}
The branch points are at $\Delta = 4 \phi_2^3 + 27 \phi_3^2 = 0$,
i.e. $z = \pm m^{3}$.
Now we consider a small perturbation of the phase $\vartheta$ away from
$\vartheta = \pi$. The resulting spectral network, determined using
the Mathematica notebook \cite{swn-plotter}, looks like
Figure \ref{fig:mass-deformed-zoom}.

\insfigscaled{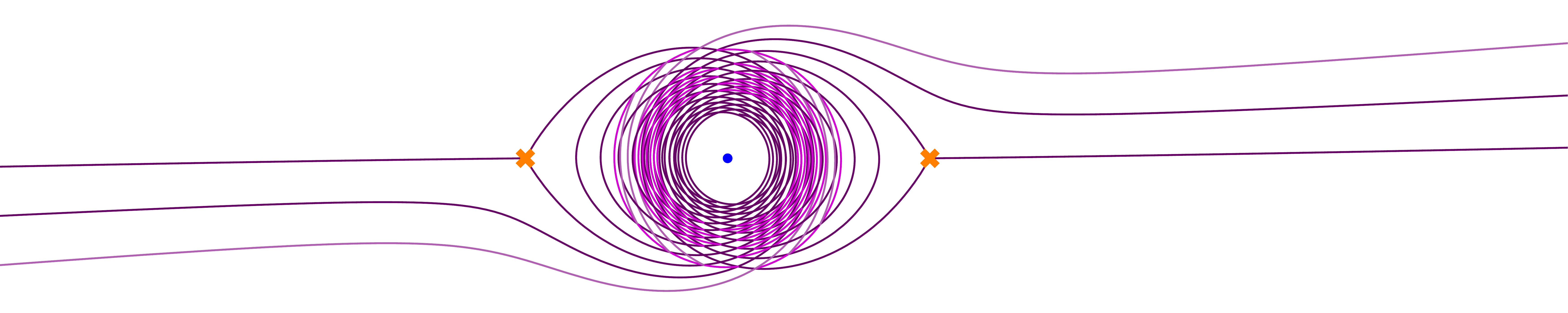}{0.22}{A spectral network
$\cW(\vartheta = \pi + 0.015)[\Lambda = 32]$, in the theory
corresponding to the differentials \eqref{eq:mass-perturbed-differentials}
with $m=1$. The
empty region around $z=0$ comes from the truncation to solitons
of mass $< \Lambda$. The colors of walls indicate their genealogy: lighter
walls are born from the intersections of darker ones.}

Since we have perturbed to a generic phase,
each wall in Figure \ref{fig:mass-deformed-zoom} supports only one soliton charge.
The three walls headed to the left are carrying the ${\bf \overline 3}$
while the three headed to the right are carrying the ${\bf 3}$.
In the massless limit $m \to 0$, the flavor symmetry
is restored, and the three walls on each side merge
into a single wall carrying the whole multiplet.
Moreover, in this limit all the complexity in the middle of
Figure \ref{fig:mass-deformed-zoom} gets squashed into a single
point, the massless puncture; thus the picture reduces
to Figure \ref{fig:decoupled-puncture}.

A similar analysis was carried out in
\cite{Maruyoshi:2013fwa} Section 4.3,
which considered a full puncture
in a different $\cS[A_2]$ theory but reached the same
conclusion that this puncture emits two triplets of
walls (under (4.20) of that paper.)

The basic phenomenon that walls of the spectral network
can emanate from a puncture, and carry
multiplets of the flavor symmetry at the puncture,
is not limited to the case of theory $T_3$;
we expect to find it generically in theories of class $\cS$
with nonabelian flavor symmetries.
(See also \cite{Gaiotto:2009hg}
Section 3.2.9 and \cite{Maruyoshi:2013fwa} for
related discussion.)
It would be interesting to explore
more systematically the landscape of possible punctures and what kind
of walls they emit.
(In the case of theories of class $\cS[A_1]$ the answer is simple: there
is only one type of regular puncture, and when its $SU(2)$
flavor symmetry is unbroken it emits a single wall,
carrying the representation ${\bf 2}$. This can
be verified by an analysis very similar to what we have done here.)

\subsection{The actual solutions}

In the actual situation of our interest, we
do not have $\nu_n = 0$: rather, from
Figure \ref{fig:circle-network-resolved} and Figure
\ref{fig:local-puncture-resolved} it follows that each
$\nu_n$ for a given puncture is related to a $\tau_n$ for a
neighboring puncture, since each ``incoming'' wall is also
``outgoing'' from a neighboring puncture. Reintroducing the index
$\ell \in \{1,2,3\}$ to keep track of the punctures
(recall $z_1 = 1, z_2 = \omega, z_3 = \omega^2$)
this gives the relations
\begin{equation} \label{eq:transport}
  \nu_{1,\ell} = x \tau_{2,\ell+1}, \quad \nu_{2,\ell} = x \tau_{1,\ell-1},
\end{equation}
where the factors $x = X_{\gamma_1}$ arise
from the ``continuation'' discussed in \S\ref{sec:counting-solitons}.
Using \eqref{eq:transport} to eliminate the $\nu_{n,\ell}$ from
\eqref{eq:puncture-traffic-rule},
we obtain a system of $6$ algebraic equations for
$6$ unknown functions $\tau_{n,\ell}(x,\{A_\ell\},\{B_\ell\})$:
\begin{equation} \label{eq:full-constraint-equations}
\tau_{1,\ell} = \frac{B_\ell + A_\ell x \tau_{2,\ell+1} + x^2 \tau_{2,\ell+1}^2 + x \tau_{1,\ell-1}}{1 - x^2 \tau_{2,\ell+1} \tau_{1,\ell-1}}, \qquad \tau_{2,\ell} = \frac{A_\ell + B_\ell x \tau_{1,\ell-1} + x^2 \tau_{1,\ell-1}^2 + x \tau_{2,\ell+1}}{1 - x^2 \tau_{2,\ell+1} \tau_{1,\ell-1}}.
\end{equation}

The full solutions to these equations seem to be rather unwieldy.
If we specialize for a moment to the case $g=1$, i.e. all $A_\ell = B_\ell = 3$,
then we obtain a simple solution which can be written
explicitly: all $\tau_{n,\ell}$ are equal and given by
\begin{equation} \label{eq:conformal-solution}
  \tau_{n,\ell} = \frac{1 - x - \sqrt{1 - 14x + x^2}}{2x} = 3 + 12x + 84x^2 + 732x^3 + 7140x^4 + \cdots
\end{equation}
Note in particular the leading $3$, which matches the $3$ light
solitons from \S\ref{sec:decoupled-limit}. All the other terms
represent heavier solitons.

\eqref{eq:conformal-solution}
is not the only solution to \eqref{eq:full-constraint-equations}
with $g=1$, but it
is the only one which has a small-$x$ expansion given by a series in
nonnegative powers of $x$, and thus it is the one which gives the
actual soliton spectrum.

\section{Computing the bulk BPS states} \label{sec:computing-bps}

After this detour to compute BPS soliton spectra on the surface
defects $\bbS_z$, we return to the question we really wanted to
answer: what is the spectrum of BPS particles
of charge $n\gamma_1$ in the four-dimensional theory $T_3$?

\subsection{The flavorless spectrum} \label{sec:flavorless}

Let us begin by computing the BPS indices at $g=1$: this
corresponds to forgetting their transformation under flavor symmetry.
We follow the recipe laid out in
Sections 6.3 and 6.4 of \cite{Gaiotto2012}.
(This recipe was derived using the ``halo picture'' which expresses
how jumps of the spectrum of 2d-4d BPS states are controlled
by the spectrum of pure 4d BPS states;
thus the $\Omega(\gamma)$ are determined
indirectly from the jump of the $\bF_\wp$ when $\vartheta$ crosses
the critical phase $\vartheta_\gamma$. For our present purposes, it will
be enough to know what the recipe is, without delving into
its derivation.)

Let $p$ denote
one of the three walls of the unresolved spectral network
(Figure \ref{fig:circle-network}).
According to \cite{Gaiotto2012}, we must
consider a product which combines contributions from the
solitons on the two constituent walls after resolving:
\begin{equation}
Q(p) = 1 + \tau \nu.
\end{equation}
(In this case the $Q(p)$ for all three $p$ are the same.)
Then we must decompose this product in the form\footnote{The passage
from $Q(p)$ to the coefficients $\alpha_n(p)$ is an example of
a ``plethystic logarithm,'' as found in \cite{MR1601666}; these
appear in many counting problems in gauge theory, as discussed
in \cite{Feng:2007ur}.}
\begin{equation}
Q(p) = \prod_{n=1}^\infty (1 - (-x)^n)^{\alpha_n(p)}.
\end{equation}
Using \eqref{eq:transport}, \eqref{eq:conformal-solution} to determine
$\tau$, $\nu$ we have
\begin{align} \label{eq:Qp}
  Q(p) &= 1 + x\left(\frac{1 - x - \sqrt{1 - 14x + x^2}}{2x}\right)^2 \\
   &= 1 + x(3 + 12x + 84x^2 + 732x^3 + 7140x^4 + \cdots)^2 \\
    &= 1 + 9x + 72x^2 + 648x^3 + 6408x^4 + 67464x^5 + \cdots \\
    &= (1 + x)^9 (1 - x^2)^{-36} (1 + x^3)^{240} (1 - x^4)^{-2160} (1 - x^5)^{21600} \cdots
\end{align}
from which we read off
\begin{equation} \label{eq:alpha-n}
(\alpha_n(p))_{n=1}^\infty = 9, -36, 240, -2160, 21600, \dots
\end{equation}

Now, the recipe of \cite{Gaiotto2012} says that
we can compute the second helicity supertrace
$\Omega(n\gamma)$ from the $\alpha_n(p)$ as follows.
Each double wall
$p$ lifts to a chain $p_\Sigma$ on $\Sigma$. We are to compute the
cycle
\begin{equation} \label{eq:l-recipe}
  L(n\gamma) = \sum_p \alpha_n(p) p_\Sigma.
\end{equation}
The homology class $[L(n\gamma)]$ is a multiple
of $\gamma$, and then
\begin{equation} \label{eq:omega-recipe}
  \Omega(n\gamma) = [L(n\gamma)] / (n\gamma).
\end{equation}
In our situation we are taking $\gamma = \gamma_1$,
each $p_\Sigma$ is separately closed and has
$[p_\Sigma] = \gamma_1$, and all three $\alpha_n(p)$ are equal,
so \eqref{eq:l-recipe}, \eqref{eq:omega-recipe}
collapse to the simple formula
\begin{equation} \label{eq:omega-from-alpha}
  \Omega(n \gamma_1) = \frac{3}{n} \alpha_n.
\end{equation}
Using \eqref{eq:alpha-n}, this gives
\begin{equation} \label{eq:bare-bps-counts}
  (\Omega(n\gamma_1))_{n=1}^\infty = 27, -54, 240, -1620, 12960, \dots
\end{equation}
These are our first BPS counts.

\subsection{The flavorful spectrum} \label{sec:flavorful-spectrum}

The result \eqref{eq:bare-bps-counts} is
encouraging: $\Omega(\gamma_1) = 27$ looks good for
a theory which is supposed to have flavor symmetry $F \simeq E_6$,
since the BPS states should be in a representation of $F$, and
the smallest nontrivial representations of $E_6$ have dimension $27$!
We might similarly guess that $\Omega(2\gamma_1) = -54$ means the states
of charge $2\gamma_1$ are in two
$27$-dimensional representations, and that $\Omega(3\gamma_1) = 240$
means three copies of the $78$-dimensional adjoint plus six
copies of the $1$-dimensional trivial representation.

As we go to larger $n$ it gets increasingly difficult to guess
the correct $E_6$ representations underlying $\Omega(n\gamma_1)$.
Fortunately we do not have to guess; we just need to
compute the BPS indices at general $g \in F_\cS$,
instead of at $g=1$. We write these indices as $\bom(\gamma_1)$.
Since $F_\cS$ and $F$ have the same rank,
knowing the transformation of the BPS states under
$F_\cS$ is sufficient to determine the full $F$ representation content.

When $g \neq 1$ we have not found an exact closed-form
expression for the $\tau_{n,\ell}$. As a matter of principle,
though, there is no problem in using
\eqref{eq:full-constraint-equations}, together with the
assumption that the small-$x$ limit is finite,
to determine $\tau_{n,\ell}$ to any finite order in
the $x$ expansion. At each order we get some polynomial in
the variables $A_\ell, B_\ell$.
For example, expanding to order $x$ gives
\begin{align}
\tau_{1,\ell} &= B_\ell + x(A_\ell A_{\ell+1} + B_{\ell-1}) + \cdots \\
\tau_{2,\ell} &= A_\ell + x(B_\ell B_{\ell-1} + A_{\ell+1}) + \cdots
\end{align}
As above we can then compute for each wall $p_\ell$
\begin{align}
  Q(p_\ell) &= 1 + \tau_{1,\ell} \nu_{1,\ell} \\
  &= 1 + x \tau_{1,\ell} \tau_{2,\ell+1} \\
  &= 1 + x (B_\ell A_{\ell + 1}) + x^2 (B_\ell^2 B_{\ell+1} + B_{\ell} A_{\ell+2} + A_\ell A^2_{\ell+1} + A_{\ell+1} B_{\ell-1} ) + \cdots
\end{align}
The next step is to expand each $Q(p)$ as a product, of the form
\begin{equation}
  Q(p) = \prod_{n=1}^\infty \prod_{\lambda \in \Lambda} (1 - (-x)^n \lambda)^{\alpha_{n,\lambda}(p)}
\end{equation}
where $\Lambda \simeq \Z^6$ denotes the character lattice of $F_\cS$.
We collect these into characters
\begin{equation}
	\bal_{n}(p) = \sum_{\lambda \in \Lambda} \alpha_{n,\lambda}(p) \lambda,
\end{equation}
and then generalizing \eqref{eq:l-recipe} we define
\begin{equation} \label{eq:bl-recipe}
	\bL(n \gamma) = \sum_p \bal_{n}(p) p_\Sigma.
\end{equation}
$\bL(n \gamma)$ is a ``character-valued $1$-cycle'' on $\Sigma$, i.e.
a $1$-cycle whose coefficients are characters of $F_\cS$ instead of
integers. $\bL(n \gamma)$ is necessarily a multiple of
$[\gamma]$, and the BPS index we are after is the coefficient:
\begin{equation} \label{eq:bom-rule}
	\bom(n \gamma) = [\bL(n\gamma)] / (n \gamma).
\end{equation}
Here we are taking $\gamma = \gamma_1$, and (as in the flavorless case
above) all $p_\Sigma$ are separately closed
and have $[p_\Sigma] = \gamma_1$. Then \eqref{eq:bom-rule} reduces to
\begin{equation} \label{eq:bom-recipe}
  \bom(n \gamma_1) = \frac{1}{n} \sum_{\ell=1}^3 \bal_n(p_\ell)
\end{equation}
generalizing \eqref{eq:omega-recipe}.

For example, we get in this way
\begin{equation} \label{eq:omega-character-1}
  \bom(\gamma_1) = \sum_{\ell=1}^3 \bal_1(p_\ell) = B_1 A_2 + B_2 A_3 + B_3 A_1.
\end{equation}
If we substitute $A_\ell = B_\ell = 3$ as before, we recover $\Omega(\gamma_1) = 27$.
More generally, from \eqref{eq:omega-character-1} we see that
$\bom(\gamma_1)$ is the character of a specific representation of
$F_\cS \simeq SU(3)^3$:
\begin{equation} \label{eq:27-decomp}
 \bom(\gamma_1) = ({\bf \overline{3}}, {\bf 3}, {\bf 1}) + ({\bf 1}, {\bf \overline{3}}, {\bf 3}) + ({\bf 3}, {\bf 1}, {\bf \overline{3}}).
\end{equation}
Since the flavor symmetry is enhanced from $F_\cS$ to
$F \simeq E_6$, we expect that
this representation should arise by decomposing some representation
of $E_6$, and indeed this is the case: \eqref{eq:27-decomp} matches
the decomposition of the irreducible representation
$\bf{\overline{27}}$.
(Our conventions for $E_6$ representations are given in
Appendix \ref{app:e6-reps}.)
We summarize this by writing:
\begin{equation} \label{eq:omega-character-1-e6}
  \bom(\gamma_1) = \bf{\overline{27}}.
\end{equation}
The formula \eqref{eq:omega-character-1-e6}
does not quite determine the spectrum of BPS particles with charge $\gamma_1$,
because of the usual possibility of cancellations in the index. The simplest possibility
would be that the spectrum consists of BPS hypermultiplets transforming in
the representation $\bf{\overline{27}}$.

A similar but longer computation leads to the result
\begin{equation} \label{eq:omega-character-2-e6}
  \bom(2\gamma_1) = -2 \times {\mathbf{27}},
\end{equation}
i.e. $-2$ times the character of the representation
${\mathbf {27}}$. Recalling that a BPS vector multiplet contributes $-2$ to $\Omega$,
the simplest possibility is that the spectrum of BPS particles of charge
$2\gamma_1$ consists
of BPS vector multiplets in the $\bf{27}$.

Continuing in this way we obtain at the next few orders
\begin{align}
  \bom(3\gamma_1) &= 3 \times {\mathbf{78}} + 6 \times {\mathbf{1}}, \label{eq:omega-character-3-e6} \\
  \bom(4\gamma_1) &= -4 \times {\mathbf {\overline {351}}} -8 \times {\mathbf {\overline {27}}}. \label{eq:omega-character-4-e6}
\end{align}

\subsection{Spin purity} \label{sec:spin-purity}

Looking at the data \eqref{eq:omega-character-1-e6}, \eqref{eq:omega-character-2-e6},
\eqref{eq:omega-character-3-e6}, \eqref{eq:omega-character-4-e6}, a
surprising phenomenon emerges: in $\bom(n\gamma_1)$ all the multiplicities
are positive integer multiples of $(-1)^{n+1} n$. In other words,
if for primitive $\gamma$ we define the reduced index
\begin{equation}
  \bored(n\gamma) = \frac{\bom(n\gamma)}{(-1)^{n+1} n},
\end{equation}
then what we have seen is that $\bored(n\gamma_1)$ is the character of
an \ti{actual} (not virtual) representation of $F \simeq E_6$.
Below we will see that this is also true up to $n \le 7$,
and we will see the same phenomenon
for several other primitive charges $\gamma$.

It would be very interesting to
understand why this is the case.
One attractive possibility is that there
is a kind of \ti{spin purity} in this theory:
all of the BPS particles of charge $n \gamma$ are in multiplets with
spin $\frac{n}{2}$. Each such multiplet contributes $(-1)^{n+1} n$ to $\bom(n\gamma)$,
so if spin purity indeed occurs, then we can interpret $\bored(n\gamma)$
simply as the count of spin-$\frac{n}{2}$ multiplets.

We note that spin purity does \ti{not} occur in arbitrary $\N=2$
theories: for example, it was shown in \cite{Galakhov2013} that
in the $\N=2$ supersymmetric pure $SU(3)$ Yang-Mills
theory, there is a point on the Coulomb branch where
the spectrum of BPS particles with fixed charge involves multiplets
of many different spins.

On the other hand, spin purity does occur
in the superconformal $\N=2$ supersymmetric $SU(2)$ Yang-Mills with $4$ hypermultiplet
flavors, on its Coulomb branch. Indeed, in that theory, along each ray
in the electromagnetic charge lattice, we have a primitive charge $\gamma$,
and the BPS spectrum consists of $8$ hypermultiplets of charge $\gamma$
plus $1$ vector multiplet of charge $2\gamma$. This example is much simpler than theory $T_3$, since it involves no BPS particles with
spin $>1$, and correspondingly no BPS particles of charge $n\gamma$
for $n > 2$.

With all this in mind, we can formulate a hypothesis: perhaps spin purity occurs
in every $\N=2$ superconformal theory on its Coulomb branch.
A weaker hypothesis would be that
spin purity occurs in every $\N=2$ superconformal theory for which
the Coulomb branch is $1$-dimensional.

\subsection{Multiplicities at higher charge} \label{sec:multiplicities-n0}

With computer assistance we computed $\bom(n\gamma_1)$ up to $n=7$,
with the following results:
\begin{center}
  \begin{tabular}{c|l}
  $n$ & $\bored(n\gamma_1)$  \\ \hline
  $1$ & $\overline{\mathbf {27}}$ \\
  $2$ & ${\mathbf {27}}$  \\
  $3$ & ${\mathbf {78}} + 2 \times {\mathbf 1}$ \\
  $4$ & ${\mathbf {\overline {351}}} + 2 \times {\mathbf {\overline {27}}}$ \\
  $5$ & ${\mathbf{1728} + 2 \times \mathbf{351} + 6 \times \mathbf{27}}$  \\
  $6$ & ${\mathbf{5824} + \mathbf{2430} + 2 \times \mathbf{2925} + 6 \times \mathbf{650} + 13 \times \mathbf{78} + 16 \times \mathbf{1}}$ \\
  $7$ & ${\overline{\mathbf{19305}}} + 3 \times {\overline{\mathbf{17550}}} + 6 \times {\overline{\mathbf{7371}}} + 13 \times {\overline{\mathbf{1728}}} + 12 \times {\overline{\mathbf{351'}}} + 29 \times {\overline{\mathbf{351}}} + 44 \times {\overline{\mathbf{27}}}$ \\
  \end{tabular}
\end{center}

With our current algorithms we were not able to go higher than $n = 7$
while keeping all the flavor information.
If we discard the flavor information, though,
we can easily use the results of \S\ref{sec:flavorless}
to compute up to $n = 200$; the first few results are:
\begin{center}
  \begin{tabular}{c|l}
  $n$ & $\ored(n\gamma_1)$  \\ \hline
  $1$ & $27$ \\
  $2$ & $27$  \\
  $3$ & $80$ \\
  $4$ & $405$ \\
  $5$ & $2592$  \\
  $6$ & $19034$ \\
  $7$ & $154224$ \\
  $8$ & $1344357$ \\
  $9$ & $12387408$ \\
  $10$ & $119234916$ \\
  $11$ & $1188951696$ \\
  $12$ & $12206381574$ \\
  $13$ & $128421415008$ \\
  $14$ & $1379545102782$
  \end{tabular}
\end{center}

We close this section with two remarks about these numbers:
\begin{itemize}

\item It is interesting to compare these results with
those of Section 5.3 of \cite{Huang2013}, where BPS degeneracies
are given for a \ti{five-dimensional} theory obtained by compactifying
$M$-theory on the cone over a del Pezzo surface $dP_6$. Upon $S^1$
compactification, this theory should reduce to the theory $T_3$ considered
here. The counts given in \cite{Huang2013}
are nonnegative integers depending on a single electric charge $n$
with $1 \le n \le 7$,
and two spins $j_L, j_R \in \half \Z$. If we simply
sum up those counts over $j_L$ and $j_R$, i.e. compute the
total number of spin multiplets, the result agrees with the
$\ored(n \gamma_1)$ computed here,
for $1 \le n \le 6$. For $n=7$, however, there is a mismatch:
summing up the degeneracies given in the last table of Section 5.3 in
\cite{Huang2013} gives $156438$, while the table above gives $154224$.
Moreover, this mismatch appears to persist for all $n \ge 7$.\footnote{We thank
the authors of \cite{Huang2013} for providing some numerical data
for $n > 7$.}

The agreement for $1 \le n \le 6$ looks unlikely to be
a coincidence, but we have not understood it:
Why is the total number of spin multiplets the
right thing to compare? Why does a mismatch appear at $n \ge 7$?
Relations between 4d and 5d BPS states have been proposed before
in the literature; see particularly \cite{Gaiotto:2005gf,Dijkgraaf:2006um}.
Those papers concern gravitational theories rather than pure field theories, but perhaps some relative of their constructions can
explain the agreement (and disagreement) we have found.

\item From a glance at the table one sees immediately that
$\Omega(n \gamma)$ grows exponentially with $n$.
The phenomenon of exponential growth of BPS spectra in
sufficiently complicated $\N=2$ theories has been noted before,
e.g. in \cite{Galakhov2013,Mainiero:2016xaj,Cordova2015}.

To study this growth more quantitatively, we use a strategy
recently employed in \cite{Mainiero:2016xaj}: we note
that the function $Q(x)$ which appeared in
\eqref{eq:Qp} obeys the algebraic equation
\begin{equation} \label{eq:algebraic-eq}
	x Q^2 - (x^2 - 6x + 1) Q + (x+1)^2 = 0.
\end{equation}
In particular, the discriminant of \eqref{eq:algebraic-eq} is
\begin{equation}
	\Delta = (x^2 - 6x + 1)^2 - 4x(x+1)^2 = (x-1)^2(x^2 - 14x + 1)
\end{equation}
which vanishes at
\begin{equation}
	x_* = (7 + 4\sqrt{3})^{-1}.
\end{equation}
Meanwhile, \eqref{eq:algebraic-eq} says $Q(x)$ can
vanish only at $x = -1$.
Thus the Taylor series expansion of $\log Q(x)$
around $x=0$ is convergent
up to $\abs{x} = (7 + 4\sqrt{3})^{-1}$, and hence its
coefficients grow as $\approx (7 + 4 \sqrt3)^n$.
As explained in \cite{Mainiero:2016xaj},
this is the same as the growth of the coefficients $\Omega(n\gamma_1)$
of the plethystic logarithm; moreover
the methods of \cite{Mainiero:2016xaj} allow
us to determine the subleading power-law behavior:\footnote{We
thank Tom Mainiero for explaining this to us.}
\begin{equation} \label{eq:omega-asymptotics}
	\abs{\Omega(n \gamma_1)} \sim c n^{- \frac52} (7 + 4 \sqrt3)^n,
\end{equation}
for some constant $c$.
This indeed matches well with the data.

\end{itemize}

\section{Other charges} \label{sec:other-charges}

So far we have discussed BPS counts $\Omega(\gamma)$
and $\bom(\gamma)$ where $\gamma$ is a multiple
of $\gamma_1$.
All of these BPS counts were computed using
the single spectral network $\cW(\vartheta=\pi)$.
More generally, we can study BPS states with any charge $\gamma \in \Gamma_\gauge$,
at the cost of having to consider more intricate spectral networks.
In this section we briefly discuss these more general charges.

\subsection{Charges and phases}

Fix a charge
\begin{equation}
\gamma_{[p,q]} = p \gamma_1 + q \gamma_2 \in \Gamma_\gauge.
\end{equation}
Then using \eqref{eq:periods} we have
\begin{equation}
 Z_{\gamma_{[p,q]}} = (p + \omega^2 q) M.
\end{equation}
Thus BPS states of charge $\gamma_{[p,q]}$ have mass
\begin{equation}
	M_{\gamma_{[p,q]}} = \abs{Z_{\gamma_{[p,q]}}} = \abs{p + \omega^2 q} M = M \sqrt{p^2 + q^2 - pq}.
\end{equation}
To study BPS states of charge $\gamma_{[p,q]}$, we need to draw
the spectral network at
\begin{equation}\label{eq:phasepq}
\vartheta_{[p,q]} = \arg (-Z_{\gamma_{[p,q]}})
\end{equation}
i.e.
\begin{equation}
\tan \vartheta_{[p,q]} = \frac{\sqrt{3}q}{q-2p}, \qquad \vartheta_{[p,q]} \in \begin{cases} (0,\pi) & \text{ if } q > 0, \\ (-\pi,0) & \text{ if } q < 0, \end{cases} \qquad \begin{tabular}{cc} $\vartheta_{[1,0]} = \pi$, & $\vartheta_{[-1,0]} = 0$, \\ $\vartheta_{[1,2]} = \frac{\pi}{2}$, & $\vartheta_{[-1,-2]} = - \frac{\pi}{2}$. \end{tabular}
\end{equation}

\subsection{Symmetries}

The networks
\begin{equation}
  \cW(\vartheta_{[p,q]}), \qquad \cW(\vartheta_{[-p,-q]})
\end{equation}
differ only by reversal of the sheet labels on each wall, and their phases
differ by $\pi$.
It follows that $\Omega(\gamma_{[p,q]}) = \Omega(\gamma_{[-p,-q]})$.
This kind of charge-conjugation symmetry is a general feature of all $\N=2$ theories.

More nontrivially, the residual $\Z_3$ symmetry on the Coulomb branch of theory $T_3$ is
also reflected in
a symmetry between spectral networks:
\begin{equation}
  \cW(\vartheta_{[p,q]}), \qquad \cW(\vartheta_{[q-p,-p]}), \qquad \cW(\vartheta_{[-q,p-q]})
\end{equation}
differ only by cyclic permutations of the
sheet labels $123$, and their phases $\vartheta$ differ by multiples of $2 \pi / 3$.
This also implies a corresponding symmetry of the BPS counts,
$\Omega(\gamma_{[p,q]}) = \Omega(\gamma_{[q-p,p]}) = \Omega(\gamma_{[-q,p-q]})$.

\subsection{Some concrete networks}

Combining these two symmetries we see in particular that the network
$\cW(\vartheta_{[1,1]})$ is identical to $\cW(\vartheta_{[1,0]})$,
up to changing the labels $ij$ on the walls.
To get a really new example we thus consider $\cW(\vartheta_{[1,2]})$,
shown on the left in Figure~\ref{fig:star-and-232-networks}.
This network has one qualitatively new feature compared to
Figure \ref{fig:circle-network}: it includes two joints where six walls meet, one at $z=0$ and one at $z = \infty$. On the right in Figure \ref{fig:star-and-232-networks} is
the next simplest network, $\cW(\vartheta_{[1,3]})$.

\insfigscaled{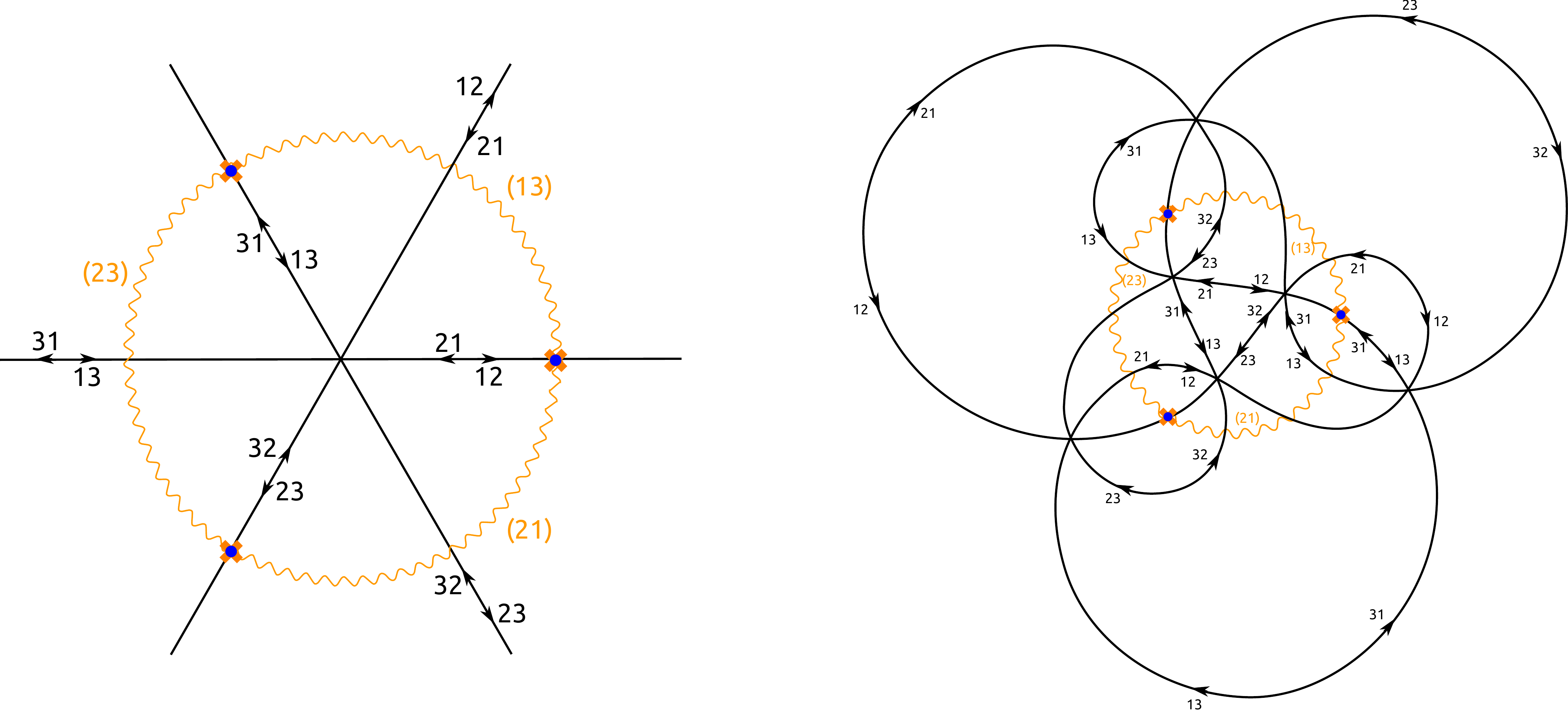}{0.35}{Left: the spectral network $\cW(\vartheta_{[1,2]})$.
Right: the spectral network $\cW(\vartheta_{[1,3]})$.
In these figures, for convenience we have chosen the branch cuts differently
than in Figure \ref{fig:circle-network}; the sheet labelings
agree inside the unit circle, but differ outside.}

As $p$ and $q$ increase, with $p$ and $q$ coprime, the networks $\cW(\vartheta_{[p,q]})$
become more intricate. $\cW(\vartheta_{[2,5]})$ is shown in
Figure~\ref{fig:25-network} below.

\insfigscaled{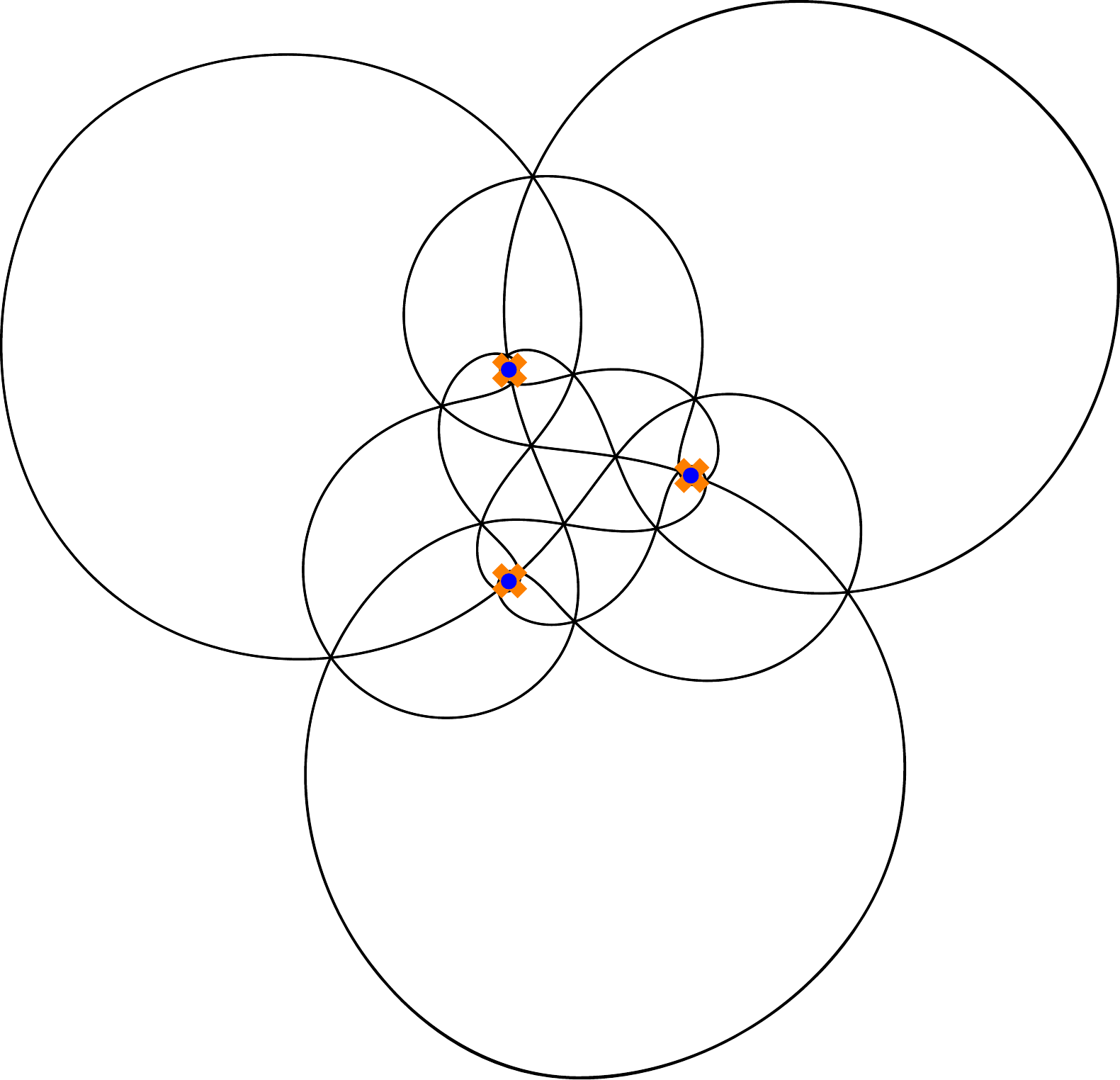}{0.5}{The spectral network $\cW(\vartheta_{[2,5]})$. All walls here are double walls, supporting solitons of types $ij$ and $ji$ simultaneously, but we do not show the labels explicitly.}

\subsection{Joints}\label{sec:joint}

One new feature of the networks $\cW(\vartheta_{[p,q]})$ is that to compute the soliton counts the constraints
\eqref{eq:puncture-traffic-rule} associated to the punctures
are no longer sufficient;
we also have to use additional local constraints associated to the
joints where six walls meet.
These constraints were described in \cite{Gaiotto2012} Appendix A.
We briefly review them here.

\insfigscaled{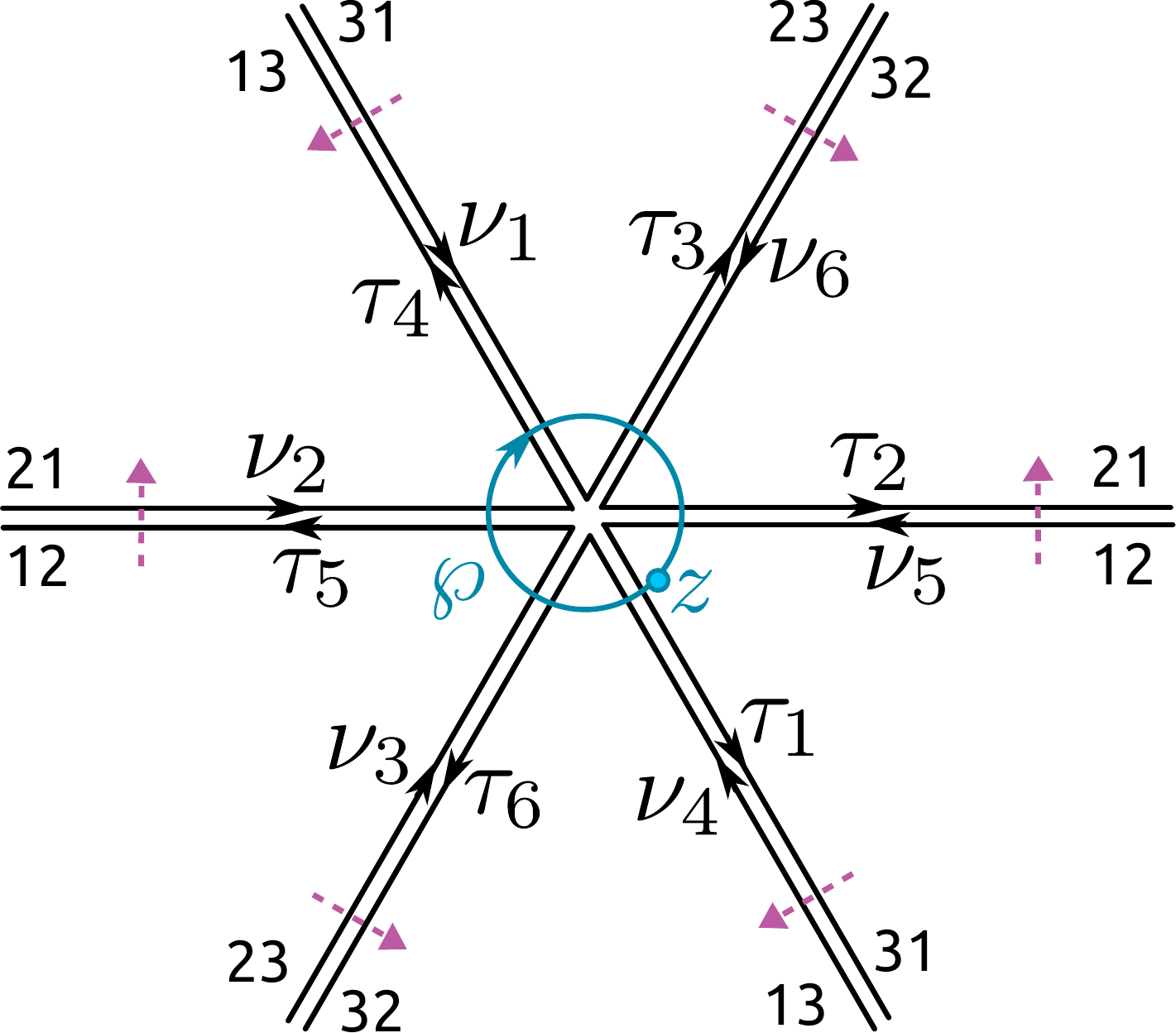}{0.4}{The local picture around a joint.}

Up to orientation-preserving diffeomorphisms in the plane, the most general joint
which can occur is shown in Figure \ref{fig:6-junction}.\footnote{In a general theory,
this statement would not make sense, since the labeling
of the sheets is arbitrary and an odd permutation
of the sheet labels produces a picture
related to Figure \ref{fig:6-junction} by a reflection. In theory $T_3$, though,
the sheets carry a natural cyclic ordering preserved by all
monodromies around branch points, and we have the relation
$\lambda^{(i+1)} = \omega \lambda^{(i)}$; this ensures that
all joints are indeed related to the one in Figure \ref{fig:6-junction}
by orientation-preserving diffeomorphism.}

As in \S\ref{sec:constrainteqns}, we have labeled the walls by symbols
$\tau_{n}$ and $\nu_{n}$ for $1 \le n \le 6$,
which we also use to represent the soliton generating functions
$\mathbf{S}_z$ on the walls.
We have also marked a loop $\wp $ on $C$ beginning and ending at the marked point $z$. We write $\wp = \wp_6 \wp_5 \cdots \wp_1$, where the $\wp_k$ are
the segments of $\wp$ running between double walls.
According to the constraints of \S\ref{sec:2d4d-constraints}, we have
\begin{multline}\label{eq:6junction}
\bF_\wp = \bF_{\wp_6} (1 - \nu_{5}) (1 - \tau_{2})\bF_{\wp_5} (1 + \nu_{6}) (1 + \tau_{3})\bF_{\wp_4} (1 - \nu_{1}) (1 - \tau_{4}) \times \\
\times \bF_{\wp_3} (1 + \nu_{2}) (1 + \tau_{5})\bF_{\wp_2} (1 - \nu_{3}) (1 - \tau_{6}) \bF_{\wp_1} (1 + \nu_{4}) (1 + \tau_{1}).
\end{multline}
Here the $\tau$'s and $\nu$'s are evaluated at the places where $\wp$ crosses the walls, and $\bF_{\wp_k}$ is the sum of three terms for the lifts of $\wp_k$ to $\Sigma$.

Now we can proceed similarly to \S\ref{sec:constrainteqns}:
express $\bF_\wp$ of \eqref{eq:6junction}
as a matrix, which by homotopy invariance
must be equal to the identity matrix;
this allows us to determine the ``outgoing'' $\tau_{n}$ in terms of the ``incoming'' $\nu_{n}$. For instance,
\begin{equation}\label{eq:solitonsjoint}
\tau_1
= \frac{\nu_1 + \nu_2 \nu_6 + \nu_1 \nu_3 \nu_6 + \nu_1 \nu_4 \nu_2 \nu_6 + \nu_1\nu_3\nu_5\nu_2\nu_6 +  \nu_1\nu_3\nu_6\nu_4\nu_2\nu_6}{1 - \nu_1\nu_3\nu_5 \nu_4 \nu_2\nu_6},
\end{equation}
where the denominator is to be expanded in a geometric series;
this gives a series expansion
with only positive signs. (Note that this would not have been
true if we had chosen different coorientations for the walls
in Figure \ref{fig:6-junction}.)

The expressions for all other $\tau_{n}$ are similar.

\subsection{Multiplicities}\label{sec:results-higher-pq}

We can now in principle compute all the BPS counts $\bom(n\gamma_{[p,q]})$ following the same algorithm described in \S\ref{sec:computing-bps}.
Unfortunately the computations required for $\cW(\vartheta_{[p,q]})$ are
generally much more expensive than those
for $\cW(\vartheta_{[1,0]})$, so we cannot compute as many BPS degeneracies.
We just describe here the results for the two networks
pictured in Figure \ref{fig:star-and-232-networks}.

For the network $\cW(\vartheta_{[1,2]})$, we find:
\begin{center}
  \begin{tabular}{c|l}
  $n$ & $\bored(n\gamma_{[1,2]})$  \\ \hline
  $1$ & $\mathbf {78} + 3 \times \mathbf{1}$ \\
   $2$ & $ {\mathbf{650}} + 2 \times \mathbf{78} + 4 \times \mathbf{1}$ \\
  $3$ & ${\mathbf{\overline {5824}}} + {\mathbf {5824}} + 2 \times \mathbf{2925} +  8 \times \mathbf{650} + 11 \times \mathbf{78} + 12 \times \mathbf{1}$ \\
  \end{tabular}
\end{center}
For the network $\cW(\vartheta_{[1,3]})$, we have only
computed one count with the
flavor information included:
\begin{center}
  \begin{tabular}{c|l}
  $n$ & $\bored(n\gamma_{[1,3]})$  \\ \hline
  $1$ & $\overline{\mathbf {351}} + 3 \times \overline{\mathbf {27}} $  \\
  \end{tabular}
\end{center}

Note that the representations of $E_6$ which appear here and in \S\ref{sec:multiplicities-n0} are constrained: indeed, on a state with
charge $\gamma_{[p,q]}$, the generator $\cC$ of
the $\Z_3$ center (see Appendix \ref{app:e6-reps} for our
conventions) acts by $\omega^{p+q}$. So, at least as far as the BPS
states of theory $T_3$ are concerned, it appears that
there is a slight mixing between
the electromagnetic symmetry and the flavor symmetry.
It would be nice to understand this on some \ti{a priori}
ground.

The flavorless BPS degeneracies for
both of these networks
can be easily computed up to $n=13$. The results are
\begin{center}
  \begin{tabular}{c|l}
  $n$ & $\ored(n\gamma_{[1,2]})$  \\ \hline
  $1$ & $81$ \\
  $2$ & $810$  \\
  $3$ & $23568$ \\
  $4$ & $1054620$ \\
  $5$ & $59272560$  \\
  $6$ & $3845869602$ \\
  $7$ & $275518046160$ \\
  $8$ & $21220796005632$ \\
  $9$ & $1727362288212480$ \\
  $10$ & $146871096341656590$ \\
  $11$ & $12936006724475199888$ \\
  $12$ & $1173014876208454094700$ \\
  $13$ & $108997909913288073225456$ \\
  \end{tabular}
\end{center}
and
\begin{center}
  \begin{tabular}{c|l}
  $n$ & $\ored(n\gamma_{[1,3]})$  \\ \hline
  $1$ & $432$ \\
  $2$ & $63126$  \\
  $3$ & $25837040$ \\
  $4$ & $15997511988$ \\
  $5$ & $12414634813584$  \\
  $6$ & $11112471629495966$ \\
  $7$ & $10976477695048905264$ \\
  $8$ & $11652623904520407820032$ \\
  $9$ & $13070660396858566472984064$ \\
  $10$ & $15312115043824353889100152626$ \\
  $11$ & $18579553424056358193512622811248$ \\
  $12$ & $23208045406405864226170364128108836$ \\
  $13$ & $29704725146725768042434236249559752976$ \\
  \end{tabular}
\end{center}

\section{A picture of the lightest states}\label{sec:geom-bps-1}

For the \ti{lightest} BPS states counted by a given
spectral network, the recipe
which we reviewed in \S\ref{sec:flavorful-spectrum}
simplifies considerably.
Indeed, let $\gamma$ be the charge of these states;
then for each double wall $p$, to compute the coefficient $\bal_1(p)$
we only need to study the lightest solitons supported on $p$:
\begin{equation}
	\bal_1(p) = \tau_{\light}(p) \nu_{\light}(p),
\end{equation}
where $\tau_{\light}, \nu_{\light}$ are the truncations
of $\tau, \nu$
to the lightest soliton charges --- i.e. we keep only
the first term in the expansion in powers of $X_\gamma$.
Then we use the rule \eqref{eq:bl-recipe} which says
\begin{equation} \label{eq:bl-light}
	\bL(\gamma) = \sum_p \bal_1(p) p_\Sigma,
\end{equation}
and as above
\begin{equation} \label{eq:bom-light}
	\bom(\gamma) = \bL(\gamma) / [\gamma].
\end{equation}

The lightest solitons are relatively easy to compute with bare hands,
and lead to simple geometric pictures
of what the BPS index $\bom(\gamma)$ is counting.
Informally speaking, we just have to count the possible ways of
``gluing together two solitons head-to-head.''
We now illustrate this in a few examples.

\subsection{Lightest states with charge \texorpdfstring{$\gamma_1 = \gamma_{[1,0]}$}{gamma1 = gamma[1,0]}}

In this section we reconsider the result
$\bom(\gamma_{1}) = \mathbf{\overline{27}}$,
which we obtained in \eqref{eq:omega-character-1-e6}.

We consider the truncated spectral networks $\cW( \vartheta_{1})[\Lambda]$. Let us start with $\Lambda \ll M$. As shown in Figure
\ref{fig:decoupled-puncture} above,
each puncture $z_\ell$ emits three light solitons in opposite directions,
carrying the flavor representations $\mathbf{3}_\ell$ and $\mathbf{\overline{3}}_\ell$. The resulting network is shown on
the left in Figure~\ref{fig:circle-network-soliton}.

\insfigscaled{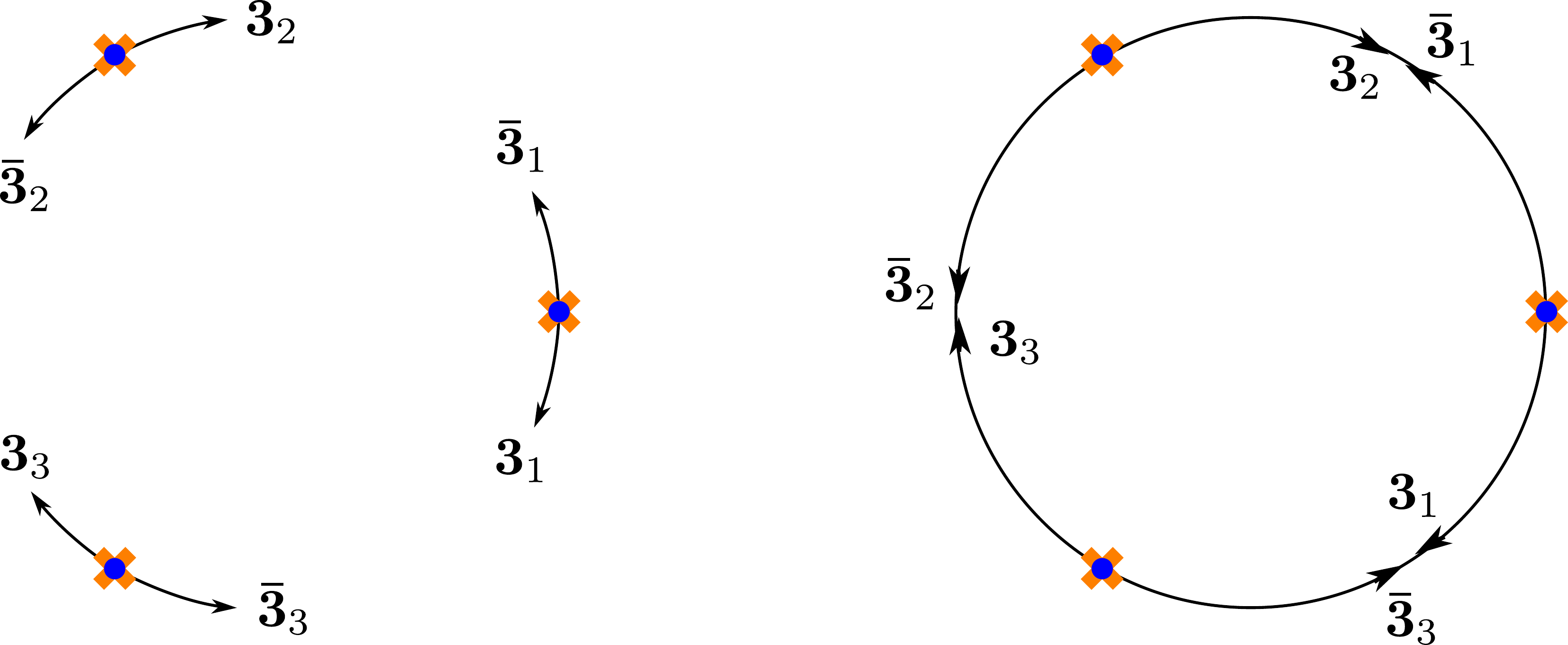}{0.3}{Solitons in the truncated network $\cW( \vartheta_{1})[\Lambda]$ for increasing values of $\Lambda$. Left: the network at $\Lambda = 0.25 M$. The wall with label $\mathbf{3}_1$ represents three light solitons in the fundamental representation of $SU(3)_1$, etc. Right: the truncated network at $\Lambda = M$. Here the walls have grown
just far enough to reach the adjacent punctures.}

As we increase $\Lambda$, the walls of the network extend, until at $\Lambda = M$ they reach the neighboring punctures. The resulting network and soliton
data are shown on the right in Figure~\ref{fig:circle-network-soliton}.
From this figure we obtain
\begin{equation}
	\bal_1(p_1) = ({\bf \overline{3}}, {\bf 3}, {\bf 1}), \quad \bal_1(p_2) = ({\bf 1}, {\bf \overline{3}}, {\bf 3}), \quad \bal_1(p_3) = ({\bf 3}, {\bf 1}, {\bf \overline{3}}).
\end{equation}
Then, according to \eqref{eq:bl-light}, $\bL(\gamma_1)$ is a sum over the three wall segments:
\begin{equation}
	\bL(\gamma_1) = ({\bf \overline{3}}, {\bf 3}, {\bf 1}) \gamma + ({\bf 1}, {\bf \overline{3}}, {\bf 3}) \gamma + ({\bf 3},  {\bf 1}, {\bf \overline{3}}) \gamma,
\end{equation}
from which we read off using \eqref{eq:bom-light}
\begin{equation}
\bom(\gamma_{1}) = (\bf \overline{3}, \bf 3, \bf 1) + (\bf 1, \bf \overline{3}, \bf 3) + (\bf 3,  \bf 1, \bf \overline{3}),
\end{equation}
i.e. $\bom(\gamma_{1}) = \bf{\overline{27}}$, recovering
\eqref{eq:omega-character-1-e6} as desired.

\subsection{Lightest states with charge \texorpdfstring{$\gamma_{[1,2]}$}{gamma1,2}}

Next, we reconsider the BPS states of charge $\gamma_{[1,2]}$.
Figure~\ref{fig:star-network-soliton-1} and~\ref{fig:star-network-soliton-2} show a few of the relevant spectral networks at increasing values of $0 < \Lambda \le M_{\gamma_{[1,2]}}$.

As above, for small $\Lambda$ each puncture emits 3 walls in each direction. These walls are shown on the left in Figure~\ref{fig:star-network-soliton-1}. At $\Lambda = \frac{1}{3} M_{\gamma_{[1,2]}}$ these walls meet each other at the two joints at $z=0,\infty$.

Let us study what happens at the joint at $z=0$. If we substitute $\nu_{12}=\nu_{23}=\nu_{31} = 0$ into the soliton rules obtained by solving
\eqref{eq:6junction}, we find
\begin{eqnarray}
 \tau_{21} = \nu_{21},  & \tau_{32} = \nu_{32}, & \tau_{13} = \nu_{13}, \label{eq:joint1}\\
   \tau_{12} = \nu_{13}\nu_{32}, & \tau_{23} = \nu_{21} \nu_{13}, & \tau_{31} = \nu_{32} \nu_{21}. \label{eq:joint2}
\end{eqnarray}

The first line~(\ref{eq:joint1}) implies that the original walls continue to extend beyond the
joint, with the same soliton degeneracies as before.
The second line says that, in addition,
there are new walls born at the joint, whose soliton degeneracies are the product of the incoming ones; so the new walls
support solitons with flavor charge $\mathbf{3} \times \mathbf{3}$.
Their mass at the joint is $\frac{2}{3} M_{\gamma_{[1,2]}}$, the sum of
the two constituent masses.
These new walls thus only show up when
$\Lambda \ge \frac{2}{3} M_{\gamma_{[1,2]}}$.

\insfigscaled{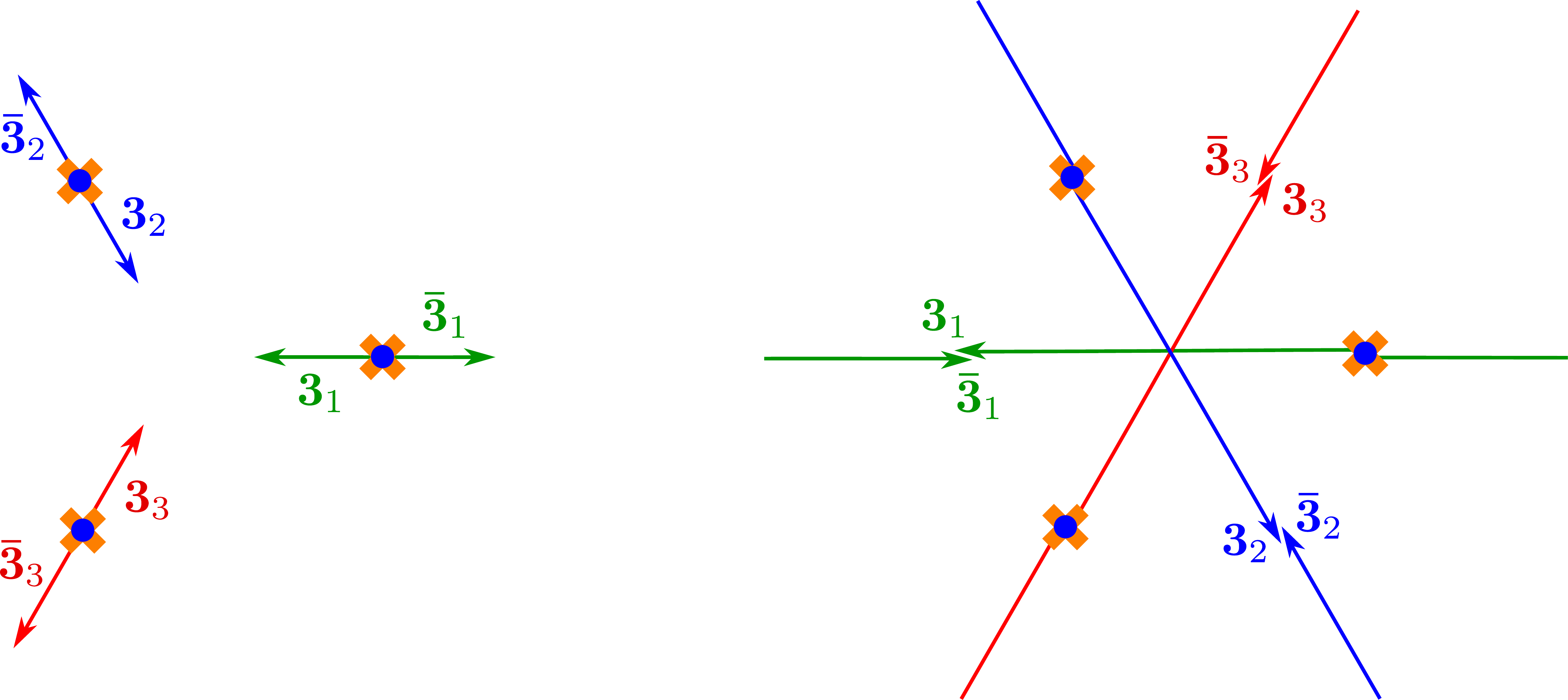}{0.25}{
Left: the truncated network $\cW(\vartheta_{[1,2]})[\Lambda]$ at $\Lambda=\frac{1}{6} M_{\gamma_{[1,2]}}$. The walls with label $\mathbf{3}$ and $\mathbf{\overline{3}}$ represent the lightest solitons emerging from the punctures. Right: the truncation at $\Lambda=\frac{1}{2} M_{\gamma_{[1,2]}}$.}

The right of Figure~\ref{fig:star-network-soliton-1} shows the truncation at  $\Lambda = \frac{1}{2} M_{\gamma_{[1,2]}}$. At this moment the original walls have extended past the joints, but the new walls have not yet been born.
The left of Figure~\ref{fig:star-network-soliton-2} shows the truncation at  $\Lambda = \frac{2}{3} M_{\gamma_{[1,2]}}$, when the new walls with label $\mathbf{3} \times \mathbf{3}$ appear in the network, emanating from the joint $z=0$. Simultaneously, the original walls with label $\overline{\mathbf{3}}$ reach the joint. Thus the new wall born from the joint at $z=0$ carries
solitons transforming in $\overline{\mathbf{3}}+\mathbf{3} \times \mathbf{3}$.
Similar comments apply to $z = \infty$.
Finally, at $\Lambda = M_{\gamma_{[1,2]}}$ all walls arrive at punctures. This is illustrated on the right of Figure~\ref{fig:star-network-soliton-2}.

\insfigscaled{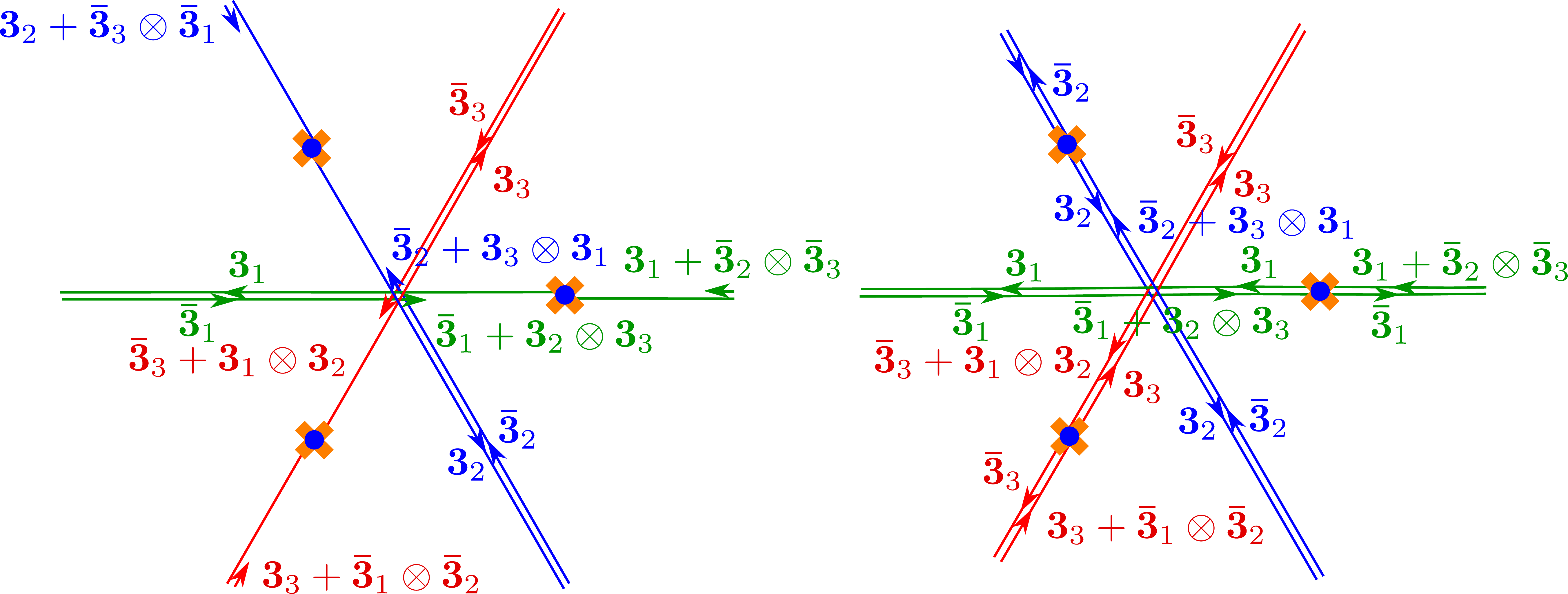}{0.25}{Left: the truncation at $\Lambda=\frac{2}{3} M_{\gamma_{[1,2]}}$. The initial walls reach the second joint and simultaneously the new walls in the tensor product representation appear. Together they combine into walls with label $\mathbf{3} + \mathbf{\overline{3}} \otimes \mathbf{\overline{3}}$. Right: the truncation at $\Lambda=M_{\gamma_{[1,2]}}$. }

Let ${p}_\ell^{a}$ be the wall segment running between the joint at $z=0$ and the puncture $z_\ell$, let ${p}_\ell^{b}$ be the wall segment between the puncture $z_\ell$ and the joint at $z=\infty$, and let $p_\ell^{c}$ be the wall segment running between $z=\infty$ and $z=0$. Then we read off
from Figure \ref{fig:star-network-soliton-2}
\begin{equation}
	\bal_1(p_\ell^a) = {\bf{3}}_\ell \otimes ({\bf{\overline{3}}}_\ell \oplus {\bf{3}}_{\ell'} \otimes{\bf{3}}_{\ell''}), \quad \bal_1(p_\ell^b) = {\bf{\overline{3}}}_\ell \otimes ({\bf{3}}_\ell \oplus {\bf{\overline{3}}}_{\ell'} \otimes{\bf{\overline{3}}}_{\ell''}), \quad \bal_1(p_\ell^c) = {\bf{3}}_\ell \otimes {\bf{\overline{3}}}_\ell,
\end{equation}
where we defined $\ell' = \ell+1$, $\ell'' = \ell+2$.

Now we use \eqref{eq:bl-light} to determine $\bL(\gamma)$. A new feature appearing in this case is that the individual lifts $p_\Sigma$
are not closed cycles: rather, we only get closed cycles once we sum up.
Nevertheless we can organize the
answer into a sum over $5$ finite string webs, as follows.
We define formal sums
\begin{equation}
w^{1}_\ell = p^a_\ell + p^b_\ell + p^c_\ell, \quad w^{2} = \sum_{\ell=1}^3 {p}^a_\ell, \quad w^{3} = \sum_{\ell=1}^3 p^b_\ell.
\end{equation}
The webs $w^{1}_\ell$ and $w^2$ are illustrated in Figure~\ref{fig:star-network-string-2}.

\insfigscaled{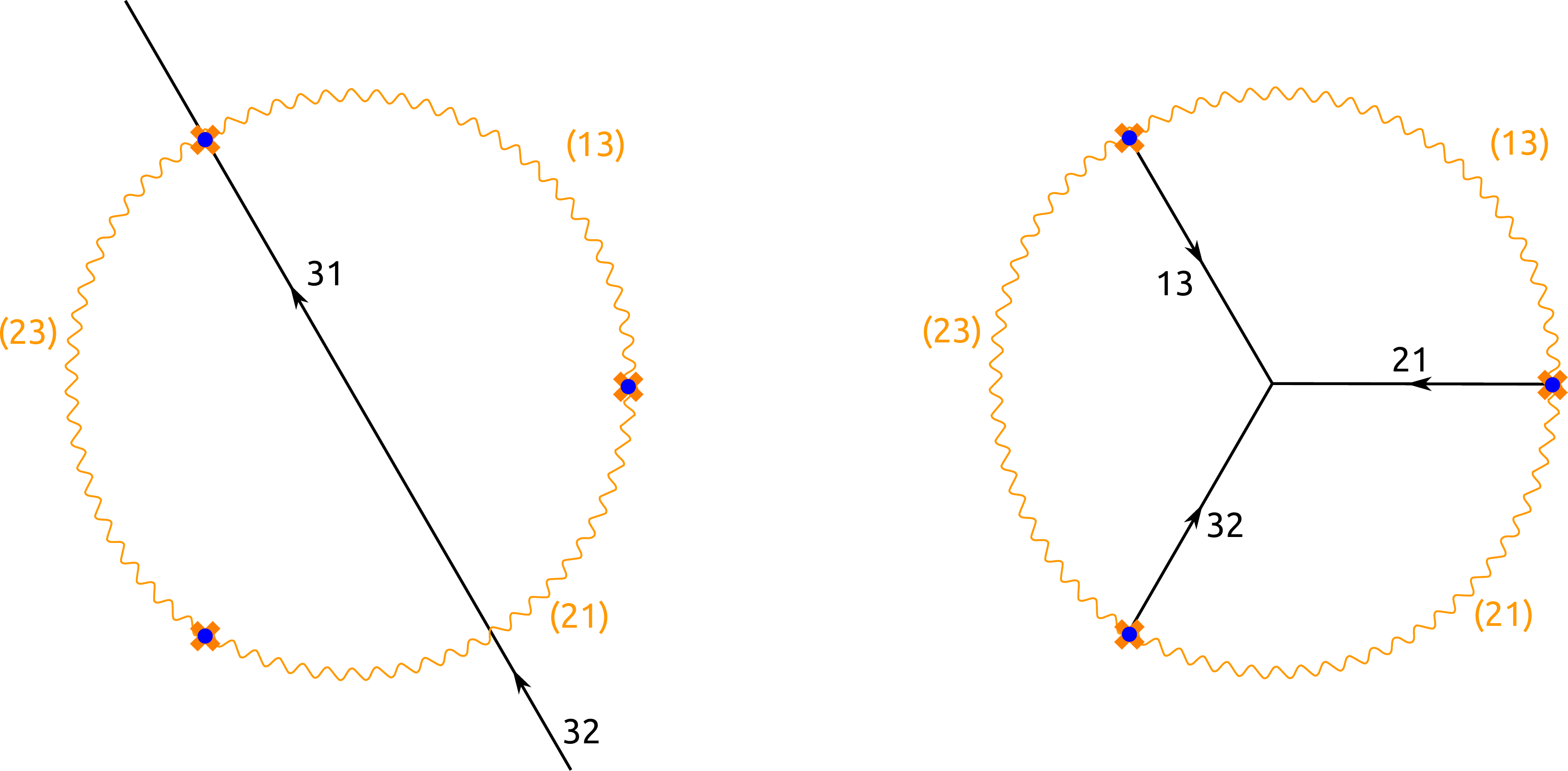}{0.33}{BPS string webs of charge $\gamma_{[1,2]}$. Left: string web $w^1_2$ with coefficient $(\bf 1, \bf{3} \otimes \bf \overline{3}, \bf 1)$. Right: string web $w^2$ with coefficient $(\bf 3, \bf 3, \bf 3)$.}

Even though each of the finite webs $w^{1}_\ell$, $w^2$ and $w^3$ has a
different topology, lifting each web to $\Sigma$ yields a 1-cycle in the
single homology class $\gamma_{[1,2]}$. The lifts of the string
webs $w^{1,2}$ and $w^2$ are shown in Figure~\ref{fig:star-network-string}.

\insfigscaled{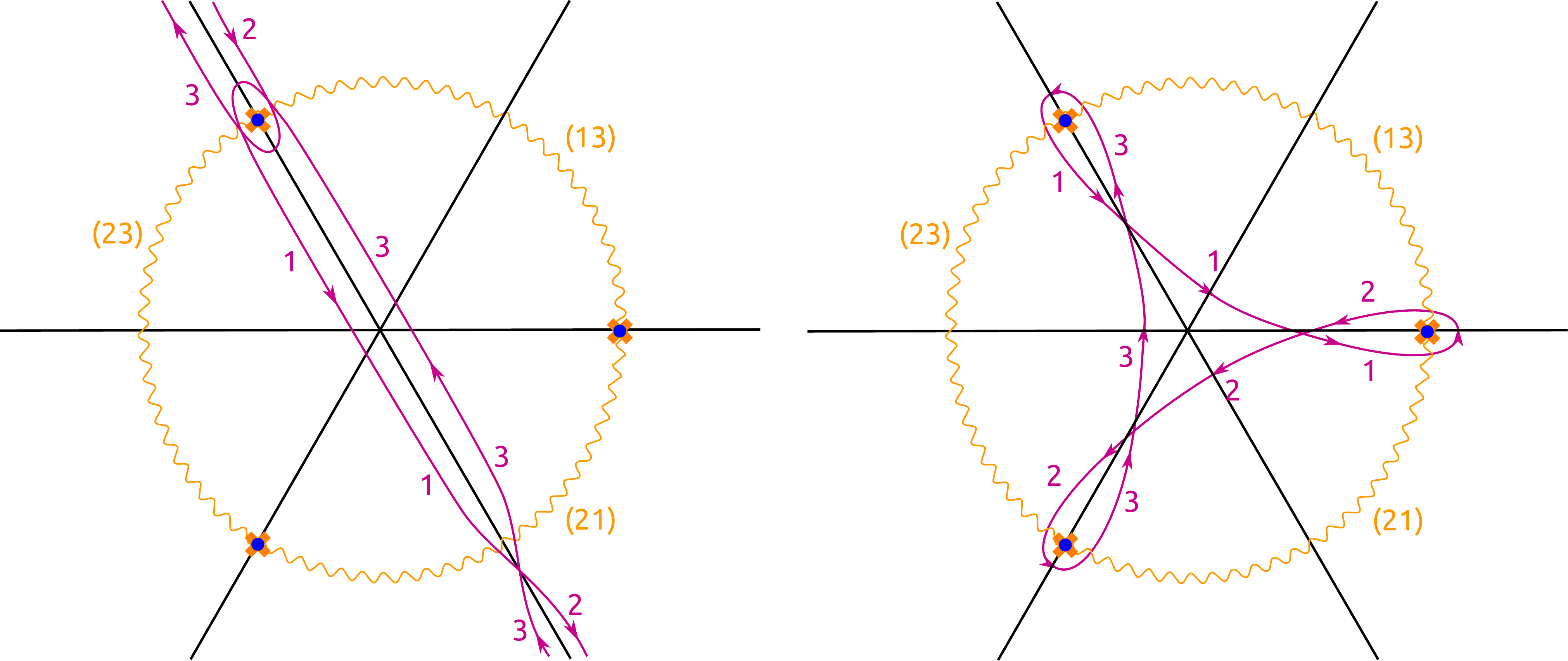}{0.33}{Lifts of the string webs in Figure~\ref{fig:star-network-string-2} to $\Sigma$.}

The cycle $\bL(\gamma)$ given by \eqref{eq:bl-light}
decomposes nicely in terms of the lifts
of these $5$ webs:
\begin{align}
\bL(\gamma) = \sum_{\ell=1}^3 ({\bf 3}_\ell \otimes {\overline{\bf 3}}_\ell) (w^{1}_{\ell})_\Sigma + ({\bf 3}, {\bf 3}, {\bf 3}) w^{2}_{\Sigma} + ( {\bf \overline{3}},  {\bf \overline{3}}, {\bf \overline{3}}) w^{3}_{\Sigma}.
\end{align}
We therefore find
\begin{align}\label{eq:starlightestBPS}
\bom(\gamma_{[1,2]}) &= (\bf{3} \otimes \bf \overline{3}, \bf 1, \bf 1) +(\bf 1,\bf{3} \otimes \bf \overline{3},\bf 1)+ (1,1,\bf{3} \otimes \bf \overline{3})+(\bf 3, \bf 3, \bf 3) + ( \bf \overline{3},  \bf \overline{3}, \bf \overline{3}) \\
&= ({\bf 3}, {\bf 3}, {\bf 3}) + ( {\bf \overline{3}},  {\bf \overline{3}}, {\bf \overline{3}})+({\bf{8}}, {\bf 1}, {\bf 1}) + ({\bf 1},{\bf{8}},{\bf 1})+ ({\bf 1},{\bf 1},{\bf{8}})+ 3 \times ({\bf 1},{\bf 1},{\bf 1}),
\end{align}
where each of the terms in the first line is directly associated
to one of the five string webs.  This is the decomposition
of the representation $\mathbf{78} + 3 \times \mathbf{1}$ of $E_6$,
matching the result we reported in \S\ref{sec:results-higher-pq}.

\subsection{Lightest states with charge \texorpdfstring{$\gamma_{[1,3]}$}{gamma1,3}}

Finally we revisit the states of charge $\gamma_{[1,3]}$. Figures~\ref{fig:232-network-soliton-1}-\ref{fig:232-network-soliton-3} illustrate the truncated networks relevant for the computation of $\bom(\gamma_{[1,3]})$.

\insfigscaled{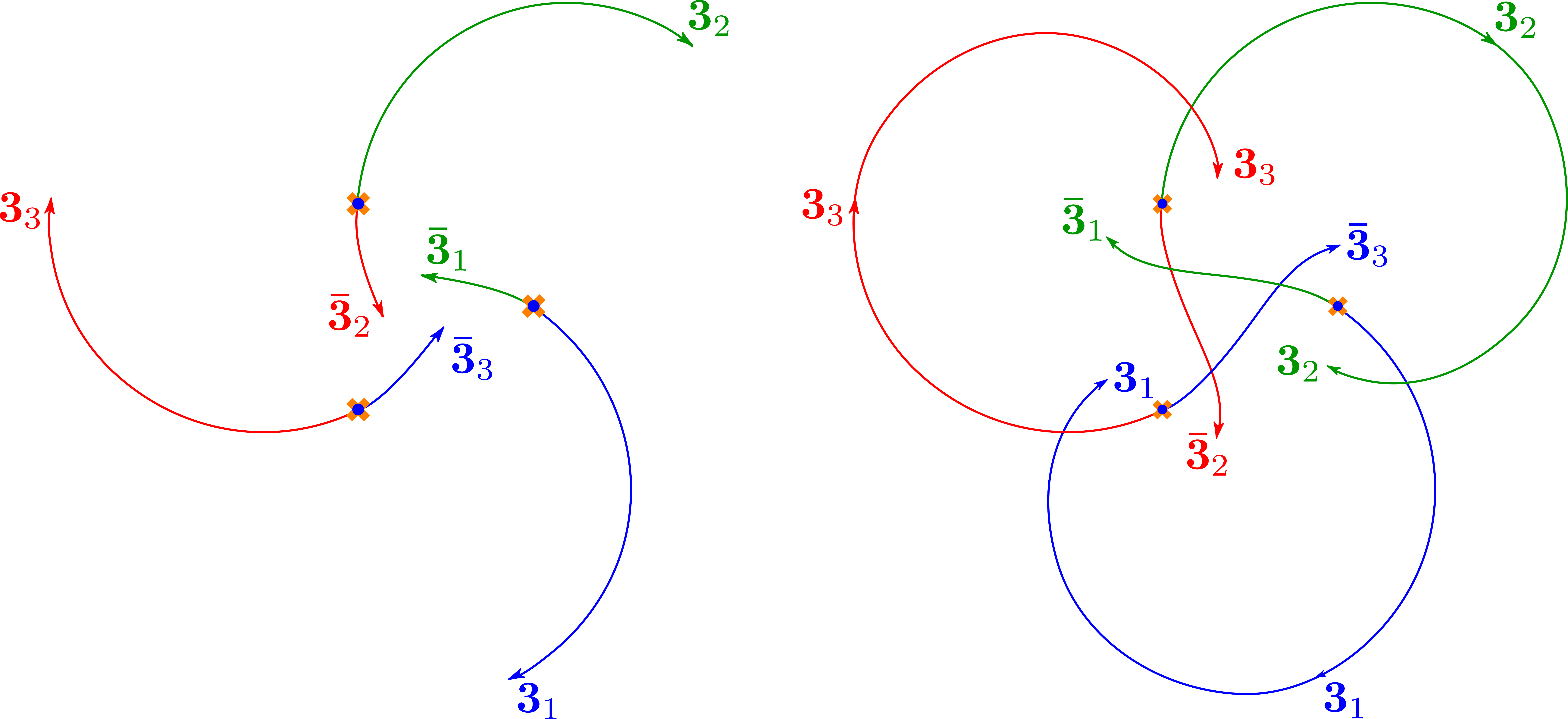}{0.15}{Left: the truncated network $\cW(\vartheta_{[1,3]})[\Lambda]$ at $\Lambda=0.2 M_{\gamma_{[1,3]}}$. At $\Lambda=\frac{2}{7} M_{\gamma_{[1,3]}}$ some of the walls meet. Right: the truncation at $\Lambda=0.32 M_{\gamma_{[1,3]}}$.}

\insfigscaled{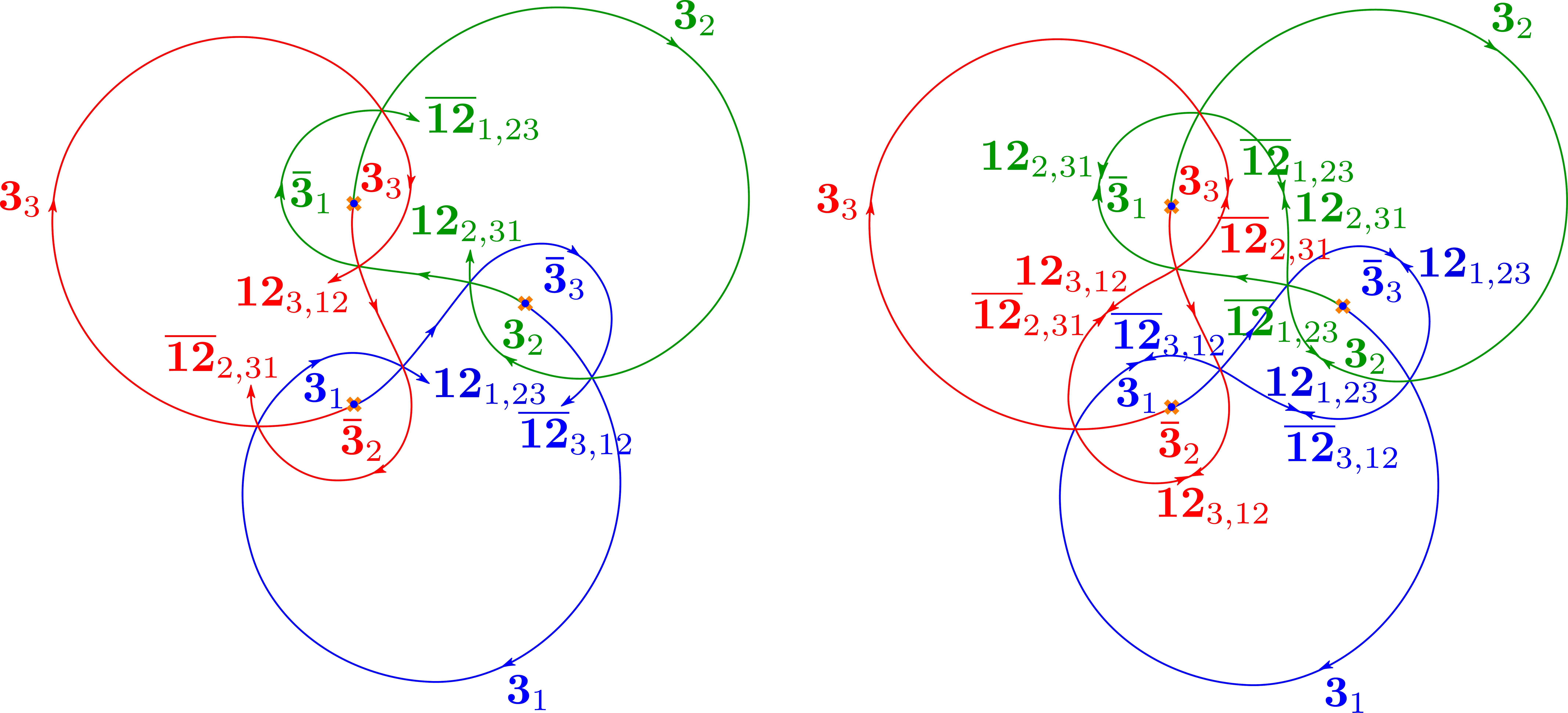}{0.15}{At $\Lambda=\frac{3}{7} M_{\gamma_{[1,3]}}$, new walls with label $\mathbf{\overline{3}} \otimes \mathbf{\overline{3}}$
are born from $3$ of the joints. At the same moment, walls with label $\mathbf{3}$
arrive at these joints. Thus the walls emerging from the joints carry the composite
label $\mathbf{12} = \mathbf{3} + \mathbf{\overline{3}} \otimes \mathbf{\overline{3}}$.  At the other $3$ joints we get similar walls with the complex conjugate
labels.
Left: the truncation at $\Lambda = 0.44 M_{\gamma_{[1,3]}}$.
At $\Lambda=\frac{4}{7} M_{\gamma_{[1,3]}}$ the walls with labels $\mathbf{12}$
and $\mathbf{\overline{12}}$ arrive at the next joint. At this joint, solitons
are generated which have $M = \frac87 M_{\gamma_{[1,3]}}$ and thus do not
contribute to the spectrum of states with mass
$M_{\gamma_{[1,3]}}$.  Right: the truncation at
$\Lambda=0.64 M_{\gamma_{[1,3]}}$.}

\insfigscaled{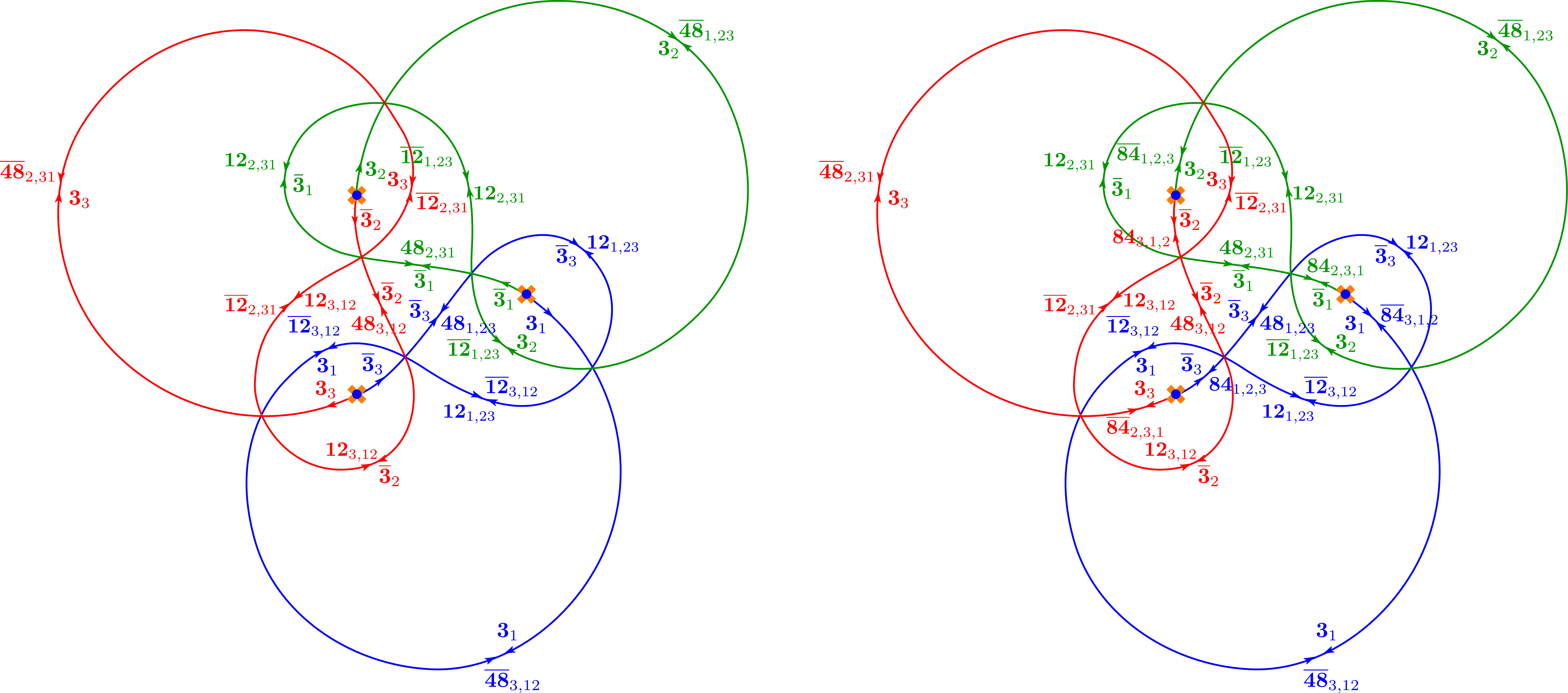}{0.157}{At $\Lambda=\frac{5}{7} M_{\gamma_{[1,3]}}$ the walls with label $\mathbf{12}$ arrive at the next joint. This generates new solitons in the representation $\mathbf{48} = \mathbf{12} + \mathbf{\overline{3}} \otimes \mathbf{\overline{12}}$. Left: the truncation at $\Lambda = 0.78 M_{\gamma_{[1,3]}}$. At $\Lambda = \frac{6}{7} M_{\gamma_{[1,3]}}$ the new walls cross the final joint. This generates solitons in the representation $\mathbf{84}=\mathbf{48}+\mathbf{\overline{3}} \otimes \mathbf{\overline{12}}$ (and complex conjugate). Right: the truncation at $\Lambda = M_{\gamma_{[1,3]}}.$
}

In the final truncation at the right in Figure~\ref{fig:232-network-soliton-3}, each colored wall consists of 7 segments. For
the wall colored red, the values of $\bal_1$ on these 7 segments are
\begin{equation} \label{eq:bal-1-red-wall}
{\bf{\overline{3}}}_2 \otimes {\bf{84}}_{3,1,2},\
{\bf{\overline{3}}}_2 \otimes {\bf{48}}_{3,12},\
{\bf{\overline{3}}}_2 \otimes {\bf{12}}_{3,12},\
{\bf{\overline{12}}}_{2,31} \otimes {\bf{12}}_{3,12},\
{\bf{\overline{12}}}_{2,31} \otimes {\bf{3}}_{3},\
{\bf{\overline{48}}}_{2,31} \otimes {\bf{3}}_{3},\
{\bf{\overline{84}}}_{2,3,1} \otimes {\bf{3}}_{3},
\end{equation}
where we defined convenient combinations of representations:
\begin{align}
\mathbf{12}_{3,12} &= \mathbf{3}_3+ \mathbf{\overline{3}}_1 \otimes \mathbf{\overline{3}}_2, \\
\mathbf{48}_{3,12} &= \mathbf{12}_{3,12} + \mathbf{\overline{3}}_3 \otimes \mathbf{\overline{12}}_{3,12}, \\
\mathbf{84}_{2,3,1} &= \mathbf{48}_{2,31} + \mathbf{\overline{3}}_2 \otimes \mathbf{\overline{12}}_{1,23}.
\end{align}
For the other two walls (blue and green), the values of $\bal_1$
are obtained from \eqref{eq:bal-1-red-wall} by a cyclic permutation
of the indices $123$.

As in the last example, we can realize the resulting
$\bL(\gamma)$ as a sum over string webs $w$.
There are $12$ such webs, some of which are illustrated in Figure~\ref{fig:232-network-string-2a} and Figure~\ref{fig:232-network-string-2b}. (The ones that are not shown can be obtained by applying rotations by multiples
of $\frac{2\pi}{3}$ to the ones shown.)

\insfigscaled{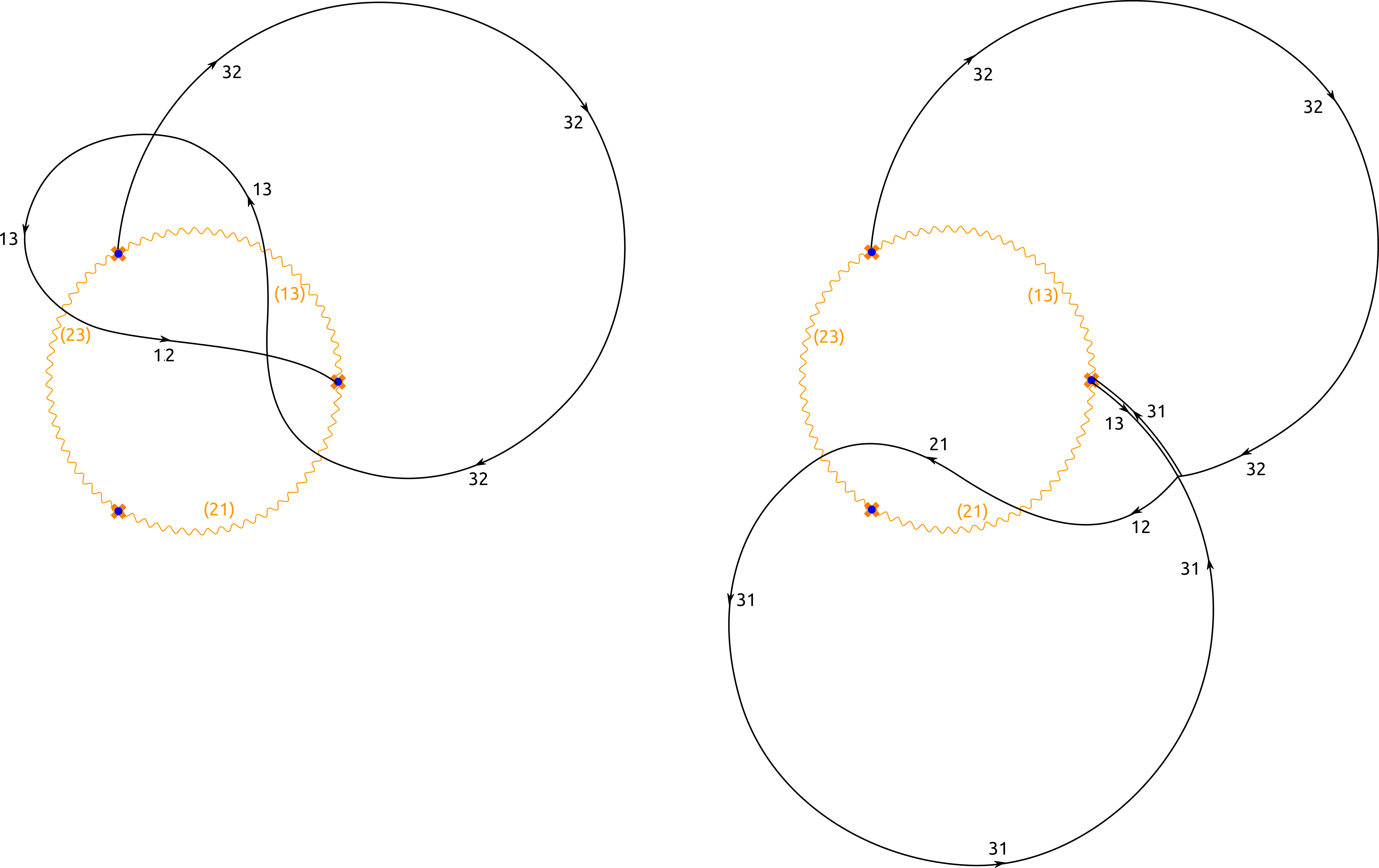}{0.2}{BPS string webs of charge $\gamma_{[1,3]}$. Left: string web with coefficient $(\mathbf{\overline{3}},\mathbf{3},\mathbf{1})$. Right: string web with coefficient $(\mathbf{3 \otimes 3},\mathbf{3},\mathbf{1})$.}

\insfigscaled{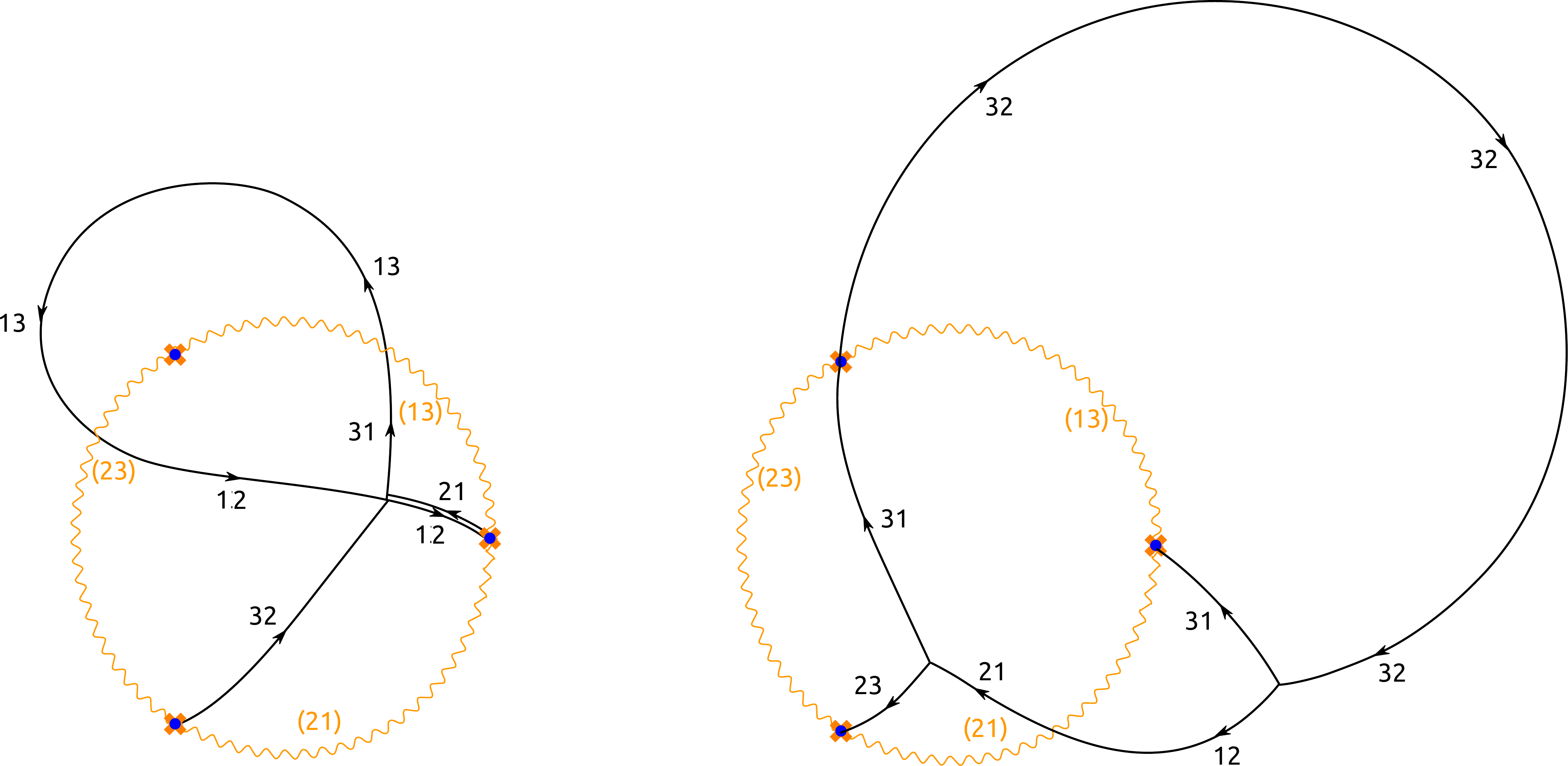}{0.2}{More BPS string webs of charge $\gamma_{[1,3]}$. Left: string web with coefficient $(\mathbf{\overline{3} \otimes \overline{3}},\mathbf{1},\mathbf{\overline{3}})$. Right: string web with coefficient $(\mathbf{3},\mathbf{3 \otimes \overline{3}},\mathbf{\overline{3}})$.}

All these webs lift to 1-cycles on $\Sigma$ in the same homology class $\gamma_{[1,3]}$.
We therefore find
\begin{align}
\bom(\gamma_{[1,3]}) =~& ({\bf 3} \otimes {\bf\overline{3}},
{\bf \overline{3}}, {\bf {3}}) + ({\bf{3}}, {\bf 3} \otimes {\bf\overline{3}},
{\bf \overline{3}}) + ({\bf \overline{3}}, {\bf {3}}, {\bf 3} \otimes {\bf \overline {3}}) + ({\bf 1}, {\bf \overline{3}}, {\bf \overline{3}} \otimes {\bf \overline{3}}) \label{eq:232lightestBPS} \\
&+ ({\bf \overline{3}} \otimes {\bf \overline{3}}, {\bf 1}, {\bf \overline{3}}) + ({\bf \overline{3}}, {\bf \overline{3}} \otimes {\bf \overline{3}}, {\bf 1}) + ({\bf {3}} \otimes {\bf {3}}, {\bf {3}}, {\bf 1})
+ ({\bf {3}}, {\bf 1}, {\bf {3}} \otimes {\bf {3}}) \notag \\
& + ({\bf 1}, {\bf {3}} \otimes {\bf {3}}, {\bf {3}}) +({\bf {3}}, {\bf 1}, {\bf \overline{3}})
+ ({\bf 1}, {\bf \overline{3}}, {\bf {3}}) + ({\bf \overline{3}}, {\bf {3}}, {\bf 1}) \notag \\
=~& ({\bf 8}, {\bf\overline{3}}, {\bf 3}) + ({\bf 3}, {\bf 8}, {\bf \overline{3}}) + ({\bf\overline{3}}, {\bf 3}, {\bf 8}) +
({\bf 1}, {\bf\overline{3}}, {\bf\overline {6}}) + ({\bf \overline {6}}, {\bf 1}, {\bf \overline {3}}) + ({\bf\overline {3}}, {\bf\overline {6}}, {\bf 1}) \label{eq:232decomp} \\
&+({\bf 6}, {\bf 3}, {\bf 1}) + ({\bf 3}, {\bf 1}, {\bf 6}) + ({\bf 1}, {\bf 6}, {\bf 3})
+4 \times ({\bf 3}, {\bf 1}, {\bf \overline{3}}) + 4 \times ({\bf 1},{\bf \overline{3}},{\bf 3}) + 4 \times ({\bf \overline {3}}, {\bf 3}, {\bf 1}), \notag
\end{align}
where each of the terms in \eqref{eq:232lightestBPS} has an interpretation as a flavor multiplet of BPS states associated to one of the $12$ webs. The result \eqref{eq:232decomp} is the decomposition of the representation $\mathbf{\overline{351}} + 3 \times \mathbf{\overline{27}}$ of $E_6$, matching what we stated in \S\ref{sec:results-higher-pq}.

\subsection{Arbitrary \texorpdfstring{$(p,q)$}{(p,q)}} \label{sec:arbitrary-pq}

Finally we can discuss more general
$\gamma_{[p,q]}$. We will show that $\Omega(\gamma_{[p,q]}) > 0$
for any $p,q$ with $(p,q)=1$, i.e.,
BPS states exist with \ti{all} primitive electromagnetic charges.

For this purpose we need some general way of understanding
what the networks $\cW(\vartheta_{[p,q]})$ look like. This turns out
to be surprisingly easy: as we now explain, we can relate
the walls of $\cW(\vartheta_{[p,q]})$ to straight-line
trajectories on an auxiliary torus. (This is similar to
a construction used in \cite{Gaiotto:2009hg}
Section 10.7 to analyze the spectrum of $\N=2$ supersymmetric $SU(2)$
Yang-Mills with $4$ hypermultiplet flavors.)

If we choose a soliton central charge
$Z_a$ for $a \in \Gamma_{ij,zz}$ as a local coordinate around $z$,
then in this coordinate the walls of
$\cW(\vartheta)$ carrying label $ij$
are just straight lines of inclination $\vartheta$ (this follows
immediately from the definition of $\cW(\vartheta)$.)
Shifting $a$ to $a+\gamma$ has the effect of shifting
the coordinate $Z_a$ by $Z_\gamma$, which by \eqref{eq:periods}
lies in the lattice
\begin{equation}
	\Xi = (\Z \oplus \omega \Z) M.
\end{equation}
We also have the $\Z_3$ action $\rho: \Gamma_{ij,zz} \to \Gamma_{i+1,j+1,zz}$, which has $Z_{\rho(a)} = \omega Z_a$.
Thus, if $a \in \Gamma_{ij,zz}$ and $a' \in \Gamma_{kl,zz}$,
$Z_a$ and $Z_{a'}$ differ by the composition of
translation by an element of $\Xi$ and multiplication
by some power of $\omega$. Said otherwise, mapping a point $z \in C$
to the collection of all soliton central charges $Z_a$ at $z$
gives a map
\begin{equation}
\phi: C \to T^2 / \Z_3, \qquad	T^2 = \C / \Xi.
\end{equation}
This map takes all three punctures of $C$ to the point
$0 \in T^2$. Lifting $\phi(\cW(\vartheta))$ from
$T^2 / \Z_3$ to $T^2$ (just taking the inverse image),
we obtain a collection
of straight lines on $T^2$, with inclinations
$\vartheta + \frac{2n\pi}{3}$.
This collection must contain at least the $3$ straight lines
emanating from $0 \in T^2$.
In Figure \ref{fig:lifted-walls} we show these lines
for $\vartheta = \vartheta_{[p,q]}$ with $(p,q) = (1,0)$, $(1,2)$, and $(1,3)$.

\insfigscaled{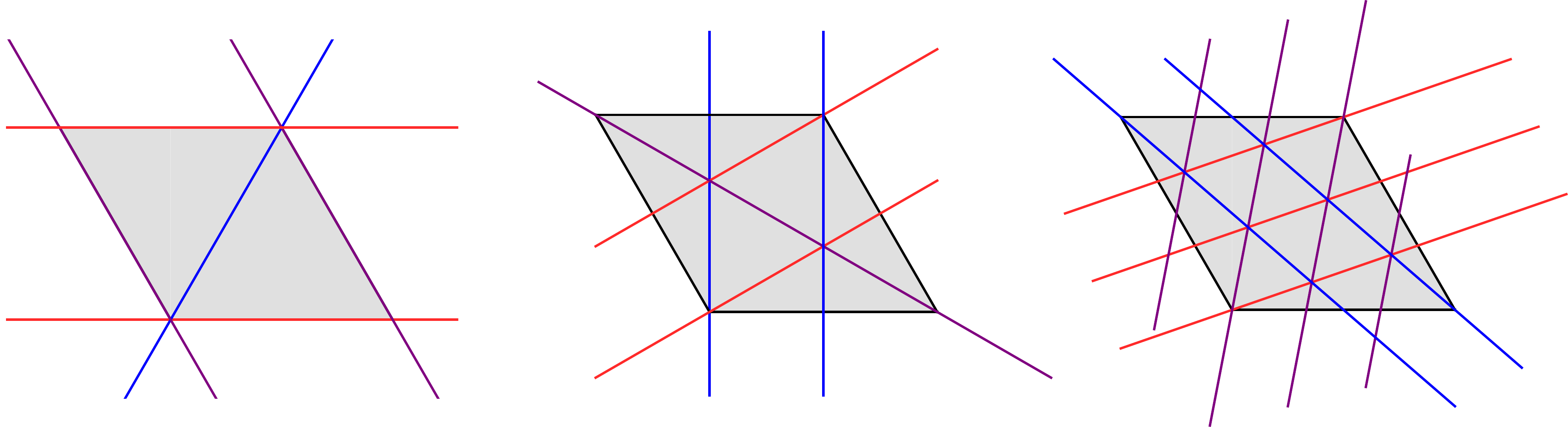}{0.42}{The walls of $\cW(\vartheta_{[p,q]})$
lifted to a fundamental domain of $T^2 = \C / \Xi$,
for $(p,q) = (1,0)$, $(1,2)$, and $(1,3)$.}

If there are any other walls of $\cW(\vartheta)$,
they must emanate from intersections between the walls
we have already drawn.
Now, any intersection between the walls on $C$ would lead to
an intersection between our lifted lines on $T^2$.
Looking at Figure \ref{fig:lifted-walls}, we see that
for $\vartheta = \vartheta_{[p,q]}$
some such intersections do exist, but
we will not obtain any new lines on $T^2$ in this way: all six possible
directions in which a new line could emanate are already populated.
We also see that, when $\vartheta = \vartheta_{[p,q]}$,
the lifted walls are all closed loops on $T^2$.
(In contrast, for all other phases $\vartheta$,
the lifted walls run around $T^2$ forever,
filling it up densely.)

Each joint on $C$ maps to an intersection
on $T^2 / \Z_3$, thus to $3$ intersections on the covering $T^2$.
On the other hand, there is also a $\Z_3$ action on $C$
by $z \mapsto \omega z$, under which $\phi$ is invariant;
joints related by this symmetry
map to the same intersection on $T^2 / \Z_3$.
There is an exceptional case if a joint occurs at
$z = 0$ or $z = \infty$: then it is a fixed point of
the $\Z_3$ action on $C$, and also its image is a fixed point of the
$\Z_3$ action on $T^2$.
Thus, in all cases the number of intersections
on $T^2$ (excluding the point $0$) is the same
as the number of six-way joints between walls in $\cW(\vartheta)$.
Comparing Figure \ref{fig:lifted-walls} with Figures
\ref{fig:circle-network}, \ref{fig:star-and-232-networks}
we see that the numbers $0$, $2$, $6$
of intersections indeed match the corresponding numbers of joints.
However, we stress that there is \ti{not} a natural
1:1 correspondence between the joints and the intersections, nor
between the walls on $C$ and the lines on $T^2$.

Nevertheless, we can read out from this picture useful
facts about the walls on $C$:

\begin{itemize}
\item Each wall, when continued far enough in either direction, ends on
a puncture.

\item The ``lengths'' of all wall segments on $C$ --- as measured by
the change in the soliton
mass as we move along the wall --- are equal. (This follows
from the fact that the segments on $T^2$ have equal length,
as visible in Figure \ref{fig:lifted-walls}.)

\item
There exists a global coorientation of all walls on $C$,
such that the local picture around each joint matches
Figure \ref{fig:6-junction}. (This coorientation corresponds to one
of the two possible $\Z_3$-invariant coorientations of the lines on $T^2$.)

\item
With this coorientation, the local picture around each puncture
matches Figure \ref{fig:local-puncture-resolved}.
(To see this, first note an invariant characterization
of the coorientation in Figure \ref{fig:local-puncture-resolved}:
the outgoing wall of type $i,i+1$ has coorientation corresponding
to going clockwise around the puncture,
and the outgoing wall of type $i+1,i$ has coorientation going
counterclockwise. On the other hand, in Figure
\ref{fig:6-junction}, the incoming walls of type $i,i+1$ have
the counterclockwise coorientation around the joint,
and the incoming walls
of type $i+1,i$ have the clockwise coorientation. These two
are compatible, as desired.)
\end{itemize}

The existence of this global coorientation, together with the fact
that all signs in the expansion of \eqref{eq:solitonsjoint} and
\eqref{eq:puncture-traffic-rule} are
positive, shows that the soliton generating functions $\tau$, $\nu$ have
all coefficients nonnegative. In particular, this is true
of $\tau_\light$ and $\nu_\light$. It follows that
(letting $\gamma = \gamma_{[p,q]}$)
\begin{equation}
	L(\gamma) = \sum_p c_p p_\Sigma
\end{equation}
where all $c_p > 0$, and the sum runs over all double walls
(moreover, since all walls end on punctures, this sum is nonempty.)
Now, we can restate \eqref{eq:omega-recipe} as
\begin{equation}
	\Omega(\gamma) = \frac{\int_{L(\gamma)} \E^{-\I \vartheta} \lambda}{\int_{\gamma} \E^{-\I \vartheta} \lambda} = \sum_{p} c_p \frac{\int_{p_\Sigma} \E^{-\I \vartheta} \lambda}{\int_\gamma \E^{-\I \vartheta} \lambda}
\end{equation}
and recalling that both $\int_{p_\Sigma} \E^{-\I \vartheta} \lambda$
and $\int_\gamma \E^{-\I \vartheta} \lambda$ are negative real numbers,
we conclude that
\begin{equation}
	\Omega(\gamma) > 0
\end{equation}
as we claimed.

\appendix

\section{\texorpdfstring{$E_6$}{E6} representations} \label{app:e6-reps}

We label the nodes of the $E_6$ Dynkin diagram by:

\begin{center}
   \begin{tikzpicture}[scale=.55]
    \foreach \x in {1,...,5}{
      \pgfmathparse{ 2*\x +3 };
      \pgfmathresult;
      \let\y\pgfmathresult;
      \fill[black] (\y , 2) circle (6pt) node[anchor=north] {};
    }
   \draw[color=black, line width=1pt] (5,2) -- (13,2);
   \draw[color=black, line width=1pt] (9,2) -- (9,4);
   \fill[black] (9,4) circle (6pt) node[anchor =east] {$6$~~};
   \draw  (5,1.4) node[color= black] {$1$};
   \foreach \x in {2,...,5}{
      \pgfmathparse{ 2*\x + 3};
      \pgfmathresult;
      \let\y\pgfmathresult;
      \draw  (\y,1.4) node[color= black] {$\x$};
    }
   \end{tikzpicture}
\end{center}

Then our names for the representations are:

\begin{center}
\begin{tabular}{c|c}
Dynkin labels & Representation \\ \hline
000000 & $\mathbf {1}$ \\
100000 & $\mathbf {27}$ \\
000010 & $\overline{\mathbf {27}}$ \\
000001 & $\mathbf {78}$ \\
010000 & $\overline{\mathbf {351}}$ \\
000100 & $\mathbf {351}$ \\
200000 & $\overline{\mathbf {351}}'$ \\
100010 & $\mathbf {650}$ \\
100001 & $\mathbf {1728}$ \\
000011 & $\overline{\mathbf {1728}}$ \\
000002 & $\mathbf {2430}$ \\
001000 & $\mathbf {2925}$ \\
110000 & $\mathbf {5824}$ \\
100100 & $\overline{\mathbf {7371}}$ \\
010001 & $\overline{\mathbf {17550}}$ \\
200001 & $\overline{\mathbf {19305}}$ \\
\end{tabular}
\end{center}

These are the names used in the Mathematica package
{\tt LieART} \cite{Feger:2012bs}, which we used to perform computations.

We let $\cC$ denote a generator of the center $\Z_3$ of the
compact simply connected form of $E_6$.
On a representation with Dynkin labels $(a_1, \dots, a_6)$,
$\cC$ acts by the cube root of unity
\begin{equation}
	\omega^{a_1 - a_2 + a_4 - a_5}.
\end{equation}
So e.g. $\cC$ acts as $\omega$ on the representations ${\mathbf{27}}$
and ${\mathbf {1728}}$,
as $\omega^{-1}$ on $\overline{\mathbf {27}}$,
and trivially on the adjoint ${\mathbf {78}}$.

Our conventions for the $SU(3) \times SU(3) \times SU(3) / \Z_3$ subgroup
of $E_6$ are fixed by specifying the decomposition of the representation
${\mathbf{27}}$:
\begin{equation}
{\mathbf{27}} = (\mathbf{3}, \mathbf {\overline{3}}, \mathbf {1}) + (\mathbf {1}, \mathbf{3}, \mathbf {\overline {3}}) + (\mathbf {\overline{3}},  \mathbf {1}, \mathbf {3}).
\end{equation}
(This does \ti{not} match the conventions of {\tt LieART}: to compare,
one needs to act by the nontrivial
outer automorphism of the first $SU(3)$ factor,
which has the effect of conjugating the
representations of that factor.)

\bibliographystyle{utphys}
\bibliography{e6-paper}

\end{document}